\pdfoutput=1 
\documentclass[%
 reprint,
 twocolumn,
superscriptaddress,
 amsmath,amssymb,
 aps,
 bbm,
]{revtex4-1}

\usepackage{graphicx}
\usepackage{dcolumn}
\usepackage{bm}
\usepackage{bbm}

\usepackage{graphicx}
\usepackage[colorlinks]{hyperref}
\AtBeginDocument{%
  \hypersetup{
    citecolor=RawSienna,
    linkcolor=Red,   
    urlcolor=Black}}
\usepackage{comment} 
\usepackage{color}
\usepackage{mathtools}
\usepackage{outlines}
\usepackage{amsmath}
\usepackage{amssymb}
\usepackage{amsfonts}
\usepackage{empheq}
\usepackage{bbm}
\usepackage{MnSymbol}
\usepackage[dvipsnames]{xcolor}
\usepackage[normalem]{ulem}

\def\magenta{\color{magenta}}

\usepackage{mathtools}

\begin{document}

\preprint{APS/123-QED}

\title{Emergence of robust memory manifolds}

\author{Tankut Can}

\altaffiliation{authors listed alphabetically; \newline correspondence:  \{tankut.can, kameshkk\}@gmail.com }

\affiliation{Department of Physics, Emory University, Atlanta, GA}

\author{Kamesh Krishnamurthy}

\altaffiliation{authors listed alphabetically; \newline correspondence:  \{tankut.can, kameshkk\}@gmail.com }
\affiliation{Joseph Henry Laboratories of Physics and PNI, Princeton University, Princeton NJ}






\date{\today}

\begin{abstract}

The ability to store continuous variables in the state of a biological system (e.g. a neural network) is critical for many behaviours. Most models for implementing such a memory manifold require hand-crafted symmetries in the interactions or precise fine-tuning of parameters. We present a general principle that we refer to as {\it frozen stabilisation} (FS), which allows a family of neural networks to self-organise to a critical state exhibiting multiple memory manifolds without parameter fine-tuning or symmetries. 
Memory manifolds arising from FS exhibit a wide range of emergent relaxational timescales and  can be used as general purpose integrators for inputs aligned with the manifold. Moreover, FS allows robust memory manifolds in small networks, and is therefore relevant to debates about implementing continuous attractors with a small number of neurons in light of recent experimental discoveries. \\

\noindent{\bf Teaser} 
Continuous memories and diverse timescales in biological systems, traditionally thought to require fine-tuning, can be robustly implemented in generic networks via the mechanism of frozen stabilization.

\end{abstract}

\maketitle


\section{Introduction}

Many biological systems are able to maintain a memory of relevant quantities in their dynamical states for durations much longer than the characteristic response time of the component parts.  For instance, humans can keep items in working memory for several minutes, even though the intrinsic timescale of neuronal responses is on the order of $10-100$ms.  This fundamental and critical memory function is particularly relevant for neural systems, where the key challenges of implementing memory in networks of neurons have been extensively studied (c.f. \cite{chaudhuri2016computational} and references therein). If we consider the memory as being stored in the collective activity of the neurons, then the problem becomes one of implementing fixed-points in the network dynamics. Storing a discrete set of items can be achieved by having a discrete set of stable fixed-points in the dynamics; the Hopfield model \cite{hopfield1982neural} is a prominent example of such a memory. 

Storing continuous variables in the dynamical state of a neural network is,  fundamentally, a more difficult challenge. 
It requires a continuum of  fixed-points forming a manifold, equipped with a  structure that can preserve the metric information of the continuous variables. Such a  manifold attractor will be locally stable in directions normal to the manifold, and {\it marginally stable} along the manifold, so that the state can move smoothly along the manifold.
A memory manifold like this can subserve critical functions such as evidence accumulation \cite{brody2016neural,gold2007neural}, path integration \cite{mcnaughton2006path,mcnaughton1996deciphering} and more generally implementing flexible cognitive maps \cite{whittington2020tolman,mcnaughton2006path}.  Constructing such a manifold in  neural networks has often required fine-tuned parameters and/or hand-crafted symmetries in the connectivity matrix (e.g. models for spatial navigation \cite{tsodyks1999attractor, tsodyks1995associative,burak2009accurate}, integrators \cite{goldman2010neural}, orientation tuning \cite{ben1995theory, seung1996brain,seung1998continuous}, but  also c.f. \cite{darshan2021learning,fisher2013modeling}). Implementing memory manifolds robustly is an open theoretical and experimental question.

Closely related to the problem of constructing memories is the problem of generating  long timescales in neural networks (\cite{bialek2012biophysics} pp. 329-350). This is motivated by the presence of a spectrum of timescales in the behaviours generated by these neural networks -- a memory function might require a single long timescale, however, other behaviours might contain a wide spectrum of timescales.  In these cases too, models for generating long timescales require some form of fine-tuning \cite{chen2020searching}. 

The difficulty of constructing memory manifolds goes beyond neural networks and applies to dynamical models of biological systems more generally. For a dynamical system to function as a continuous memory, it needs to be poised between instability and stability  -- i.e. be marginally stable. This is typically achieved at bifurcation points, thus requiring parameters to be fine-tuned, or by requiring special symmetries in the connectivity matrix, which is another form of fine-tuning. Perturbations  away from the fine-tuned operating point will destroy/destabilise the manifold.
Marginal stability without parameter fine-tuning necessarily requires the system to self-organise to a critical point, i.e. exhibit self-organised criticality  -- a highly non-trivial phenomenon \cite{jensen1998self}. Apart from specific examples of this phenomenon, more general principles to achieve this are lacking.

We present a principle that we refer to as {\it frozen stabilisation}, which
allows a family of neural networks to self-organise to a state exhibiting memory manifolds and a wide range of relaxational timescales. The principle works by  spontaneously  freezing/slowing the dynamics of a (random) subpopulation of neurons, thereby creating a static background input which serves to stabilise the remaining population. This stabilisation brings the system to rest on a manifold of marginally stable fixed-points, on which the state can move smoothly. These memory manifolds can function as general purpose neural integrators \cite{goldman2010neural}, and the emergence of the memory manifold and timescales is robust to parameter choices and permits a wide range of connectivity matrices, including ones without any symmetries/structure.



\section{ Robust memory manifolds in a simple recurrent network}
\label{sec:frozen-stab-gRNN}

We first illustrate the frozen stabilisation principle by means of a specific neural network model, and then discuss the general principle, which applies more broadly, in the next section. 

\subsection{Marginal stability over a range of parameters}

Let us consider a recurrent neural network (RNN) described by the following dynamics:
\begin{align} \label{eq:large-gRNN}
   \frac{d \mathbf{h}}{dt} = \sigma(W \mathbf{h}  ) 
    \odot \left[ - \mathbf{h} +J \phi(\mathbf{h}) \right]
\end{align}

Here, $\mathbf{h}$ is a $N-$dimensional vector representing the internal states of the neurons, $J,W$ are $N \times N$ connectivity matrices, $\sigma(x) = [1+e^{-\alpha x}]^{-1}$ is a sigmoidal non-linearity that acts element-wise and $\phi(x) = \tanh(gx)$ is the neuronal output nonlinearity with gain $g$. 
The entries of $J,W$ are sampled i.i.d from a centered Gaussian distribution with variance $1/N$. 
In \cite{krishnamurthy2020theory}, a variant of this ``gated" RNN (gRNN) was studied and shown to exhibit marginally-stable fixed points over a range of parameters, in the limit of a binary sigmoid ($\alpha \to \infty$). 

Specifically, in the marginally stable regime, the instantaneous Jacobian $\mathcal{D}$ describing the linearised dynamics, has the maximal value of the real part of the spectrum exactly equal to $0$ -- i.e. the system is poised between stability and instability. And, unlike classical RNNs which permit this behaviour only for a single, fine-tuned value of the gain $g$ \cite{sompolinsky1988chaos}, the gRNN permits this behaviour for $1<g \lessapprox 3.27$ (see Appendix \ref{app-ssRNN} for details). In this wide range of parameters, the dynamics will flow to one of the many marginally-stable fixed points, which form a manifold. Note that the matrices $(J, W)$, are completely unstructured, and the emergence of the manifold does not require any symmetries in the connectivity.

\subsection{Mechanism for marginal stability in the gRNN}
\label{subsec:frozen-stab-mech-gRNN}

In order to understand frozen stabilisation, it is instructive to study the dynamical mechanism which gives rise to the self-organised critical state in the gRNN. One might na\"{i}vely expect that in the limit $\alpha \to \infty$, the $\{\sigma \}$ terms go to zero and the $h-$variable becomes frozen, leading to fixed-points. However, with a little reflection it should be clear that at any given time the argument of the gating function $\{[W h]_i\}$ will take both positive and negative values, and thus one should expect half the $\{ \sigma_i \}$ to be zero and the other half to be 1 -- i.e. half the $\{h_i\}$ are momentarily frozen and the other half active, or free to evolve. Moreover, we know from prior work that  a ``vanilla'' RNN (vRNN) -- where $ \sigma_i =1 $ for all $i$ in eq.\ref{eq:large-gRNN}  --  exhibits chaotic  activity for $g>1$ \cite{sompolinsky1988chaos,krishnamurthy2020theory}. So, in the gRNN, if half the $\{h_i\}$ are free to evolve, what causes {\it all } the $\{h_i\}$ to be frozen -- i.e. be at a (marginally stable) fixed-point?

This is where the {\it stabilisation} in frozen stabilisation enters. As mentioned above, a snapshot in time reveals  a (random) partition of the system into two populations depending on whether $\sigma_{i}$ is $0$ (frozen) or $1$ (active). For simplicity, and without loss of generality, we can relabel neurons so that $\sigma_{i} = 1$ for $i = 1, ..., N_{\rm a}$. Let us denote the active population by $h_{i}^{\rm a} = h_{i}$ for $i = 1, ..., N_{\rm a}$, and the frozen population by $h^{\rm f}_{\mu} = h_{N_{\rm a} + \mu}$ for $\mu = 1, ... , N_{\rm f}$. In these variables, the dynamics of the network  (\ref{eq:large-gRNN}) take the form
\begin{align} \label{eq:two-pop-eom}
    \frac{d {\bf h}^{\rm a}}{dt} = - {\bf h}^{\rm a} + J^{\rm a} \phi({\bf h}^{\rm a}) + \pmb{\beta}^{\rm f}, \quad \frac{d{\bf h}^{\rm f}}{dt} = 0.
\end{align}

Here, $J^{a}$ is a $N_{a} \times N_{a}$ submatrix (upper-left block) of $J$ with entries $J^{a}_{ij} \sim \mathcal{N}(0, \mu_{\sigma}/N_{a})$, and $\mu_{\sigma} = N_{a}/N$ is the total fraction of active neurons. The first equation in eq. \ref{eq:two-pop-eom} is written to explicitly show that the interactions from the frozen population, $\beta_{i}^{f} = \sum_{j}J_{i\mu}^{f} \phi(h_{\mu}^{f})$, play the role of a {\it static} noise (or ``bias''), where $J^{\rm f}$ is a ${N_{\rm a} \times N_{\rm f}}$ submatrix of $J$ with entries $J_{i\mu}^{\rm f} \sim \mathcal{N}(0, (1-\mu_{\sigma})/N_{f})$, and $N_{\rm f} = N -N_{\rm a}$.  The interactions amongst the active population result in a time-dependent noise process  $\eta_{i}(t) = \sum_{j}J_{ij}^{\rm a} \phi(h_{j}^{\rm a}(t))$. We motivate splitting the terms into static and dynamic components below.

We know from prior work on vRNNs that  static, random external biases like $\beta_{i}^{\rm f}$ in eq. (\ref{eq:two-pop-eom}) have a stabilising effect on the dynamics, and for $ g> 1$, the dynamic noise from the recurrent interactions destabilises fixed-points, leading to chaos \cite{schuecker2018optimal, kadmon2015transition,molgedey1992suppressing}. Thus, the static and dynamic noise have opposing effects, and when the strength/variance of the static noise is above a critical value, it can overcome the dynamic noise, making the dynamics flow to a (non-zero) fixed-point. Crucially,  for a given set of external biases ${\beta_i}$ in the vRNN, there is typically only {\it one} fixed-point. 

\begin{figure}
\begin{centering}
\includegraphics[scale=0.5
]{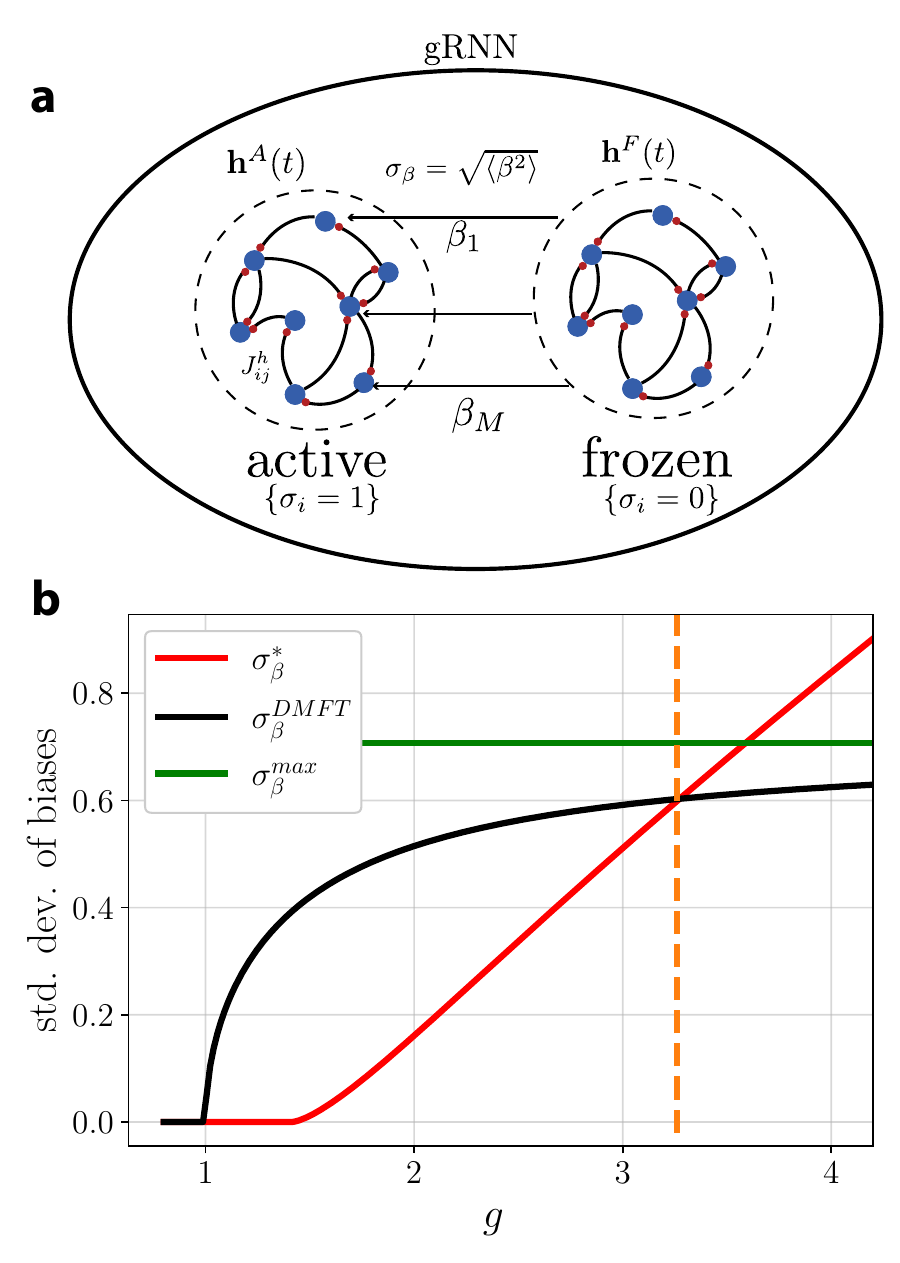}
\par\end{centering}
\caption{\label{fig:vRNN_biases} {\it Illustrating frozen stabilisation  in a RNN:} a) Schematic of the gated RNN (gRNN)  depicting the spontaneous division into frozen ($\mathbf{h}^{\rm f}$) and active ($\mathbf{h}^{\rm a}$) subpopulations. The static inputs/biases to the active population are shown as $\beta_1 \ldots \beta_M$ with variance $\sigma_{\beta}^2$. The dynamics of the active subpopulation will flow to a  fixed-point if the variance of the biases is larger than a critical value $\sigma^{*}_{\beta}(g)$.
b)  DMFT calculation of the  critical variance $\sigma^{*}_{\beta}$ required to stabilise the dynamics of the active population ( red line), and the typical variance of the biases from frozen population in the gRNN $\sigma^{DMFT}_{\beta}$ (black line). The green line shows the maximum possible value of the variance of the biases from the frozen population, when all the frozen units are saturated ($\sigma_{\beta}^{max} = 1/\sqrt{2}$).    For $1<g \lessapprox 3.27$ (till the orange line) the frozen population provides biases that can stabilise the evolving part  for typical configurations. 
Note that for $g < \sqrt{2}$, the critical bias is zero, reflecting the fact that any amount of bias will stabilise a nonzero FP.
}
\end{figure}

In the gRNN, the interactions $\beta_{i}^{f}$ from the frozen population effectively act as static ``external biases'' to the active population, with the important difference that these biases are not fixed constants but functions of dynamical variables. If the variance of interactions from the frozen population is strong enough, then we expect -- in analogy to the vRNN -- that the dynamics of the active population will be stabilised, and the whole system will flow to a fixed point. 
The critical value of the variance of biases, $\sigma_{\beta}^{*}$, that will stabilise the dynamics of the active population is calculated using MFT in Appendix \ref{app-ssRNN}, and plotted as a function of $g$ in Fig. \ref{fig:vRNN_biases}c (red line). 

We can also use MFT to calculate the {\it typical} value of the variance of the static noise/bias from the frozen population in the gRNN ( Appendix \ref{app-ssRNN}). 
This quantity, $\sigma_{\beta}^{DMFT}$, is shown in  Fig. \ref{fig:vRNN_biases}c (black line).  We require $\sigma_{\beta}^{DMFT} > \sigma_{\beta}^{*}$ in order for the frozen population to stabilise the active population, and thus make the whole system flow to a fixed point. Indeed, we see that this is the case for all $g$ that satisfy $1<g \lessapprox 3.27$ (compare red and black lines in Fig. \ref{fig:vRNN_biases}c). Thus, whenever $g$ satisfies this condition  the network self-organises to a state where the {\it frozen} population acts as random biases that {\it stabilise} the active/free-to-evolve population, thus making the whole system marginally stable. It is important to emphasise that simply having half the population  frozen ($\sigma_i=0$) will not give rise to the self-organised state -- i.e. there is no marginal stability for $g > 3.27$ (orange line in Fig.\ref{fig:vRNN_biases}c); both freezing and stabilising are important. We also emphasise that the composition of the frozen/active population are not fixed, and the ``biases'' provided by the frozen population are dynamical variables and not fixed constants. As will be shown below, this results in a self-organised state that is characterised by a continuum of connected fixed-points -- i.e. a memory manifold.

Finally, we comment on the dimensionality and the stability of the manifolds that can emerge from this model. As we show below, the local dimension $D$ of the memory manifold is given by the total number of frozen neurons, or $D = (1 - \mu_{\sigma}) N$. Moreover, the discussion above illustrates that the static  noise/bias must be sufficiently large with respect to the dynamic noise in order for stabilisation to occur. For the gRNN, this ratio scales approximately like 
\begin{align}
\frac{\sigma_{{\rm stat}}^{2}}{\sigma_{{\rm dyn}}^{2}} \sim  \frac{1-\mu_{\sigma}}{ \mu_{\sigma}}, \label{eq:noise_ratio_grnn}
\end{align}

which will be large for small $\mu_{\sigma}$. Therefore, frozen stabilisation in the gRNN will be more robust for high-dimensional manifolds (see Fig. (\ref{fig:ssRNN-phase})). It is possible also to think of $\mu_{\sigma}({\bf h})$ as a function of the position in phase space; in this case, the relation \ref{eq:noise_ratio_grnn} implies that the dynamics is attracted to regions which have small $\mu_{\sigma}$, or have a relatively higher number of frozen neurons. We further develop this phase-space perspective in discussing the global geometry of the memory manifold in Sec.\ref{subsubsec:dim-global-geometry}.


\section{Frozen stabilisation as a general principle to produce memory manifolds}
\label{sec:frozen-stab-principle}


\begin{figure}
\begin{centering}
\includegraphics[scale=0.55
]{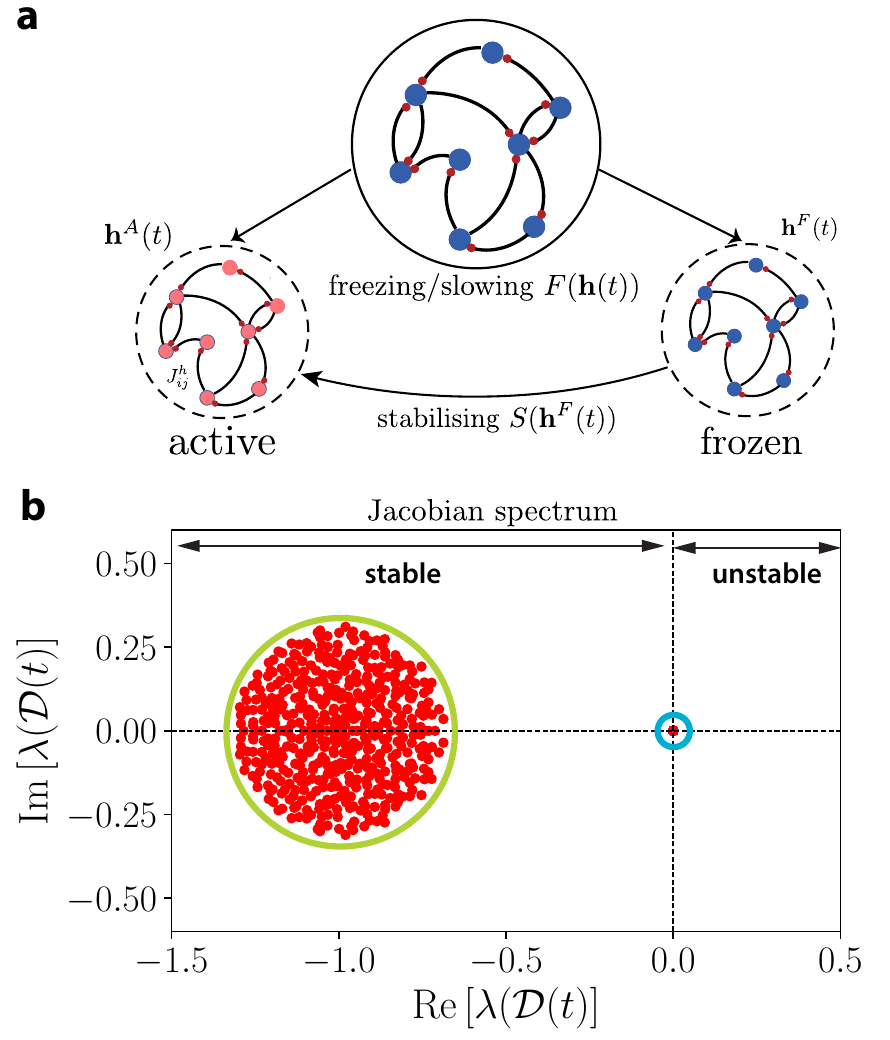}
\par\end{centering}
\caption{\label{fig:Jac_schematic} {\it Frozen stabilisation through the lens of the Jacobian:}
a) Schematic of a generic network exhibiting frozen stabilisation. A part of the system $(\mathbf{h}^{\rm f})$ gets frozen instantaneously as a function of the full state, $F(\mathbf{h}(t))$. The frozen part in turn stabilises the active/free-to-evolve part, $(\mathbf{h}^{\rm a})$,  by means of the interactions $S(\mathbf{h}^{\rm f}(t))$.
b) The Jacobian spectrum of a generic system exhibiting frozen stabilisation. Any freezing mechanism gives rise to an accumulation of eigenvalues at zero (blue circle), and if the frozen part can sufficiently stabilise the evolving part, then the remaining eigenvalues will be in the left half-plane (green circle). }
\end{figure}


The mechanism discussed in the context of the gRNN, where one part of the system that is spontaneously slowed, stabilises the freely evolving part, forms the core  of the principle of frozen stabilisation for achieving the memory manifolds. Although, we illustrated the principle with a specific model, we will see that it is more general and does not depend on the specific parameter and connectivity choices of the gRNN.  Fig. \ref{fig:Jac_schematic}, shows a schematic of a system displaying frozen stabilisation and the accompanying Jacobian spectrum. 
The main requirement for the principle is that one part of the system $(\mathbf{h}^{\rm f})$ gets instantaneously slowed/frozen depending on the state of the whole system (depicted by $F(\mathbf{h}(t))$ in Fig. \ref{fig:Jac_schematic}a). The freezing function $F$ we consider in the case of the gRNN are the switch-like binary gates with tuneable biases, $F(\mathbf{h}) = \sigma(W \mathbf{h} )$ but we aren't limited to this choice. More generally, the freezing function $F$ is a map from the continuous activity ${\bf h} = (h_{1}, ..., h_{N})$ to a bit string  $\boldsymbol{\sigma}  = (1, 0, ..., 0, 1)$, which indicates whether a particular coordinate is frozen ($\sigma = 0$) or active ($\sigma = 1$). 

When the freezing happens in the neuron basis -- instead of a rotated basis -- the dynamics can be represented in {\it full generality} using gating variables as 
\begin{align}\label{eq:gRNN_general}
    \dot{\bf h} = {\bf G}( {\bf h}, \pmb{\sigma}) \equiv \pmb{\sigma} \odot {\bf G}({\bf h}, \pmb{\sigma}).
\end{align} 
The freezing function therefore effectively produces multiplicative interactions between a {\it internally} generated field $\pmb{\sigma} = F(\mathbf{h}(t))$ and the state vector.

As a result of the subpopulation freezing, the active population will have an effective equation of motion $ \dot{\mathbf{h}}^{\rm a} = G(\mathbf{h}^{\rm a}, \mathbf{h}^{\rm f}) $, which describes {\it non-autonomous} dynamics driven by a static input ${\bf h}^{\rm f}$ - the activity of the frozen neurons. The functional form of the static drive is denoted by $S(\mathbf{h}^{\rm f})$ in Fig. \ref{fig:Jac_schematic}a. 

If one has a network which reaches a fixed point under a static drive, frozen stabilisation then suggests a {\it general strategy} to construct a network generating memory manifolds: one must enlarge the phase space, and include a  freezing function $F({\bf h})$ which spontaneously freezes a subpopulation to provide the static drive. The static input then becomes a {\it dynamical} variable rather than a fixed external variable, and thus the  the dynamics will come to rest on a manifold of fixed points

We also note that the freezing can occur at the level of {\it modes} -- i.e. a rotated basis -- instead of the neuron basis. To see this consider dynamics with of the form 
\begin{align}
    \dot{h}_{i} = \sum_{j} M_{ij}(\mathbf{h}) G_{j}(\mathbf{h})
\end{align}

The matrix $M$ can be instantaneously diagonalised as $Q^{-1} [\pmb{\lambda}] Q$, and the velocity vector transformed $\dot{\bf h}' = Q \dot{\bf h}$ and $ G' = Q G$, such that
\begin{align}
    \dot{{\bf h}}' = \pmb{\lambda} \odot {\bf G}',
\end{align}
which can also be represented in the form (\ref{eq:gRNN_general}), but with the {\it modes} gated instead of the individual neurons. If the $F$ matrix always has zero eigenvalues, then instead of frozen neurons, in this setting there will be frozen modes, and consequently frozen stabilisation. 

It is also instructive to look at frozen stabilisation through the lens of the instantaneous Jacobian describing the linearised dynamics. Fig. \ref{fig:Jac_schematic}b shows the generic Jacobian spectrum for a system displaying frozen stabilisation: {\it any} freezing mechanism will lead to an accumulation of slow modes near zero (blue circle in Fig. \ref{fig:Jac_schematic}b) and if the frozen population can stabilise the evolving population then all the remaining Jacobian eigenvalues will have real parts less than $0$ (green circle in Fig. \ref{fig:Jac_schematic}b). 
As long as the Jacobian spectrum has this general form, we will have a system that generates a memory manifold. Since the only requirement for stable eigenvalues is that they have real parts less than $0$, the principle does not require precise connectivity matrices with special symmetries or fine-tuned parameters. Indeed, the principle will work even with: a diagonal $W$; presence of correlations between the elements of $J$ and other activation functions $\phi$, such as the rectified-linear function (ReLU). We study this robustness in more detail in Sec.(\ref{sec:robustness}).

\subsection{A self-stabilised RNN using frozen stabilisation}
\label{subsec:frozen-stab-ssRNN}

Armed with the key features of frozen stabilisation, we can now embark on constructing a variety of networks which produce a memory manifold. Here we give one example of a network than can generate a memory manifold with a lower dimension and with a small number ($\sim 20$) of neurons: the self-stabilised RNN (ssRNN).  For the ssRNN, we make use of the fact that a relatively stronger static noise  induced by the frozen population will lead to more robust stabilisation. 

We incorporate a dynamical variable that modulates the variance of the frozen population, thereby ensuring robust stabilisation of lower dimensional memory manifolds as compared to the gRNN. The dynamics of the ssRNN are given by the following equations:
    \begin{align} \label{eq:ssRNN-eom}
        \partial_t \mathbf{h} = & \boldsymbol{\sigma} \odot 
          \left[-\mathbf{h} + J (\boldsymbol{\phi} \odot {\bf r} ) \right], \\ \boldsymbol{\sigma} = & \: \Theta(W \boldsymbol{\phi} + b) , \nonumber \\ {\bf r} = & \:  \gamma \left[ \mu_{\sigma}(t)(1 - \boldsymbol{\sigma}) + G(1-\mu_{\sigma}(t) ) \boldsymbol{\sigma} \right] \nonumber 
    \end{align} 
where $\Theta(\cdot)$ denotes the Heaviside function, $\mu_{\sigma}(t) = \langle \pmb{\sigma}(t) \rangle$ is the mean (instantaneous) fraction of freely evolving neurons and $G$ is a gain control parameter  chosen to be $\leq 1$. The bias $b$ controls $\mu_{\sigma}$ and thus the dimensionality of the memory manifold. 

The self-stabilising nature of the network is evident from  $\pmb{r}$ : when the size of the evolving population is larger than the frozen population then  $\pmb{r}$  dynamically boosts the  output of the frozen population and suppresses the outputs of the evolving part thereby ensuring that the frozen part can stabilise the evolving part. Consequently, the relative magnitude of the static and dynamic drives behaves as 
\begin{align}
\frac{\sigma_{\rm stat}^{2}}{\sigma_{\rm dyn}^{2}} \sim G^{2} \frac{ \mu_{\sigma}}{1-\mu_{\sigma}},\label{eq:noise_ratio_ssrnn}
\end{align}
which is the inverse of that found for the gRNN (eq. \ref{eq:noise_ratio_grnn}). From a global geometry perspective, the relation implies that for a fixed $N$, the dynamics of ssRNN is attracted to regions with larger $\mu_{\sigma}$, and thus regions which have relatively fewer frozen neurons. This is what allows the network to self-organize to exhibit stable lower-dimensional memory manifolds. Furthermore, self-stabilisation allows the ssRNN to implement such memory manifold with typically fewer neurons. In Fig. \ref{fig:ssRNN-Jacobian} we see that a ssRNN with $N=50$ has a Jacobian that shows robust frozen stabilisation.

We would like emphasise that the ssRNN is just one example of a family of networks that can exploit frozen stabilisation to generate a memory manifold. The key requirement is that the frozen population is able to stabilise the freely evolving population thus pushing the non-zero eigenvalues of the Jacobian to the left of the imaginary line, as in Fig. \ref{fig:Jac_schematic}b. As long as the Jacobian has this form, the system will generate a memory manifold. Indeed, this principle can be implemented even in a network with $N=3$, which allows us to visualise the manifold as we see below. We now proceed to study the properties of the memory manifold produced by frozen stabilisation.


\section{Properties of the memory manifold}
\label{sec:prop-manifold}

\subsection{Robustness of Frozen Stabilisation}\label{sec:robustness}

\begin{figure}
\begin{centering}
\includegraphics[scale=0.58
]{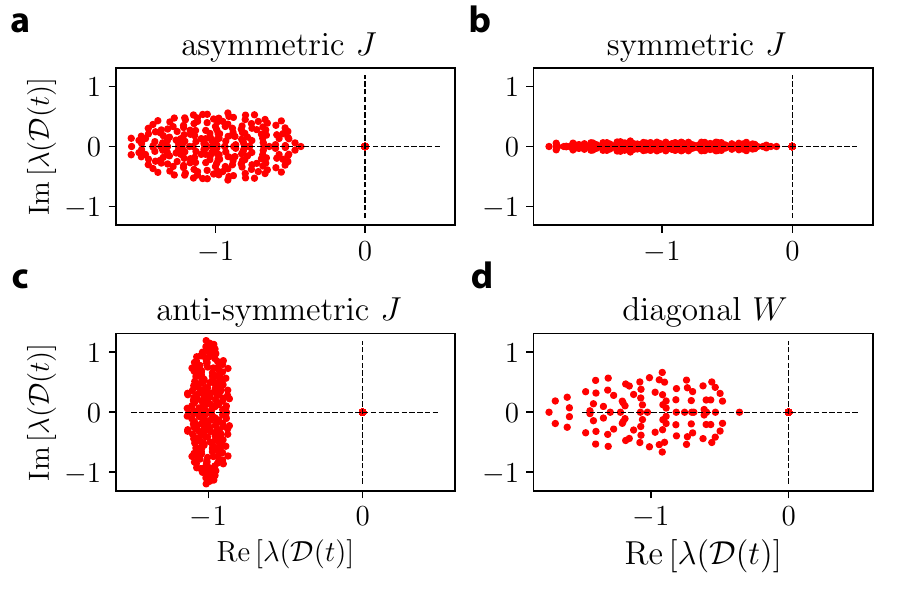}
\par\end{centering}
\caption{\label{fig:J_variations} {\it The emergence of the memory manifold is robust to choices of the connectivity matrices $J, W$} (a-c) Jacobian spectrum of a ssRNN with $J$ from an elliptic ensemble to model asymmetric (a,$\eta=0$), symmetric (b,$\eta=0.8$) or anti-symmetric (c,$\eta=-0.8$) interactions. d) Same ssRNN as (a), but with a diagonal $W$ with random $\pm 1$ on the diagonal. Parameters of the ssRNN: $N=500,g=2.0, \gamma = 1.0, G=1.0$.
}
\end{figure}

As we indicated above, frozen stabilisation and the emergence of the associated memory manifold are not slave to the details of the precise connectivity matrices, and do not require fine-tuned gain parameters. Here, we consider the effects of perturbations to parameters, and biologically relevant connectivity matrices on the emergence of the memory manifold. In networks of real neurons, a salient deviation from the completely unstructured connectivity is the over-representation of bidirectional connections suggesting a partially symmetric structure (e.g. \cite{markram1997physiology,wang2006heterogeneity}). We can study the effects of such structure on the manifold by considering matrices $J$ that can vary between fully symmetric to fully anti-symmetric. Specifically, we consider $J$ from the elliptic ensemble (\cite{sommers1988spectrum}) where $\langle (J_{ij})^2 \rangle = N^{-1} $ and $\langle J_{ij} J_{ji}  \rangle = \eta N^{-1} $, with $\eta \in [-1,1]$. The symmetric, asymmetric and anti-symmetric cases correspond to $\eta$ being $1$, $0$ and $-1$, respectively. In Fig.\ref{fig:J_variations} a-c, we see representative examples with the ssRNN wherein, other parameters being the same, introducing strong symmetry or anti-symmetry in $J$ does not affect the emergence of the memory manifold
since the non-zero eigenvalues of the Jacobian are still stable \footnote{ As an aside, the non-zero bulk of the Jacobian spectrum strongly resembles the spectrum of  $J$ from the elliptic ensemble, this is because the relevant block of the Jacobian has the form $\tilde{J} P$ where $P$ is a positive diagonal matrix and $\tilde{J}$ is from the elliptic ensemble.}.


Another important variation in the connectivity involves $W$, which governs how the freezing variable $\pmb{\sigma}$ depends on $\mathbf{h}$. In particular, when $W$ is diagonal, this would imply that the freezing variable is local -- i.e. $\sigma_{i}$ only depends on $h_i$. This choice too does not hamper the formation of the memory manifold, as is shown in an illustrative example in Fig. \ref{fig:J_variations}d. 

Thus far, we looked at choices for the connectivity matrices. Let us now briefly consider whether the memory manifold will fall apart if the connectivity matrices or the gain parameters are perturbed. For random connectivity matrices $J, W$ and large $N$,  if each $J_{ij}$ is independently perturbed by zero mean random variable with variance $\delta^{2}/N$, 
then the effect of this perturbation is similar to increasing $g \to g(1 + \delta^{2})$, and as long as the spectral edge of the non-zero eigenvalues of the Jacobian does not cross zero, the memory manifold will still exist; this will be the case when $g$ is not close to the boundaries of its permissible range. In contrast, low-rank perturbations of the form $J \to J + uu^T$ can potentially push some of the  eigenvalues to positive values, thus destabilising the slow manifold. More generally these two types of perturbations correspond to ones that globally affect  the stable bulk of the Jacobian spectrum (Fig. \ref{fig:Jac_schematic}b, green) or ones that produce outliers from the bulk, respectively.  The particular form of the detrimental perturbations will depend on the model details.

\subsection{True continuum of fixed-points}

A salient feature of the memory manifold generated by frozen stabilisation is that it is a continuum of marginally stable fixed-points. This is in contrast to previous models to generate continuous attractors, where a continuous manifold is approximated by a discrete set of $O(N)$ fixed-points, where $N$ is the system size (no. of neurons) \cite{ben1995theory,burak2009accurate,tsodyks1999attractor,si2014continuous, chaudhuri2016computational}. The distinction matters for the resolution with which the continuous variable can be stored or ``quantised''; specifically, a manifold of dimension $D$ that is approximated with $N$ fixed-points will have a resolution that scales as $N^{-1/D}$, and this can deteriorate rapidly with $D$, even for small $D$. 

To see that a network exhibiting frozen stabilisation does indeed give us a continuum of fixed-points and not a discrete set, let us consider the gRNN from before -- the argument will work with any other network exhibiting frozen stabilisation. It is sufficient to study how the network responds to infinitesimal perturbations away from a particular fixed-point. In the gRNN, denoting the frozen population by $h^{\rm f}$, and the active population by $h^{\rm a}$ (as in \ref{eq:two-pop-eom}), we find that for a neuron in the evolving population, at a fixed point the following equation must hold,
\begin{align}
  0 =  - h_{i}^{\rm a} + \sum_{j= 1}^{N_{\rm a}}J_{i j}^{\rm a}\phi(h^{\rm a }_{j}) + \sum_{\mu = 1}^{N_{\rm f}}J_{i\mu}^{\rm f} \phi(h^{\rm f}_{\mu}) .
\end{align}
 If we now apply an infinitesimal perturbation $h^{\rm f} \to h^{\rm f} + \delta h^{\rm f}$, then as long as the evolving population can still accommodate a fixed-point by adjusting infinitesimally as $h^{\rm a} \to h^{\rm a} + \delta h^{\rm a}$, the manifold will be continuous. This is possible as long as the following equation has a solution:
\begin{align} \label{eq:continuity-cond}
  \left( - \mathbbm{I} + J^{\rm a}[\phi'({\bf h}^{\rm a}) ]\right) \delta h^{\rm a} = -J^{\rm f} [\phi'({\bf h}^{\rm f}) ] \delta h^{\rm f}.
\end{align}
Here, $[\mathbf{x}]$ denotes a diagonal matrix with the vector $\mathbf{x}$ on the diagonal. There is a solution to eq. \ref{eq:continuity-cond} if the matrix in parentheses on the LHS is invertible. Since the leading edge of the spectrum of this matrix is at $-1 + \frac{1}{2} \langle \phi'^{2}\rangle_{\rm a}$ -- for large $N$ -- invertibility is guaranteed as long as $g$ satisfies the condition for marginal stability (eq. \ref{eq:large-gRNN-MS}). Thus, so long as the network exhibits frozen stabilisation, there will exist a continuous manifold of fixed-points. The same argument can be easily extended to link the stability of the Jacobian to the continuity of the manifold for other networks, like the ssRNN in Fig. \ref{fig:ssRNN-Jacobian}, exhibiting frozen stabilisation. The fact that frozen stabilisation allows networks with a small number of neurons to exhibit continuous manifolds is potentially relevant to questions about accurately storing continuous variables in networks with a few neurons in the  nervous systems of insects \cite{turner2020neuroanatomical,kim2017ring}.

\subsection{Geometry of the memory manifold}

Having demonstrated the local continuity of the memory manifold, we now study some of its geometric properties. We first discuss how the memory manifold consists of a union of attractive ``sheets'', then we provide the condition required for the stability of the sheets and show that each sheet supports relaxational dynamics with a wide range of timescales. Finally, we calculate the metric and curvature tensors for the manifold, which among other things tell us how changes in inputs are related to changes in the state of the memory manifold.

\subsubsection{Dimensionality and global geometry of the manifold}
\label{subsubsec:dim-global-geometry}

 The first point to address is the dimension of the manifold. 
Without loss of generality, we may assume we are in a region far from any boundaries, characterised by a particular bit string $\pmb{\sigma} = (0, 1,0, 0,1, ..., 0)$, where some $N_{\rm a}$ coordinates are active, and the remaining $N_{\rm f}$ coordinates are frozen.
The frozen population $(\mathbf{h}^{\rm f})$ provides a natural set of coordinates on the memory manifold, which locally looks like $\mathbbm{R}^{N_{\rm f}}$. This can  be understood by considering the structure of the left- and right-eigenvectors of the Jacobian for the general form of the dynamics (eq. \ref{eq:gRNN_general}). 
The Jacobian $\mathcal{D}$ evaluated at a fixed point far from a manifold boundary is a block structured matrix which can be written 
\begin{align}
    \mathcal{D} = [ \pmb{\sigma}] \,  \frac{\partial {\bf G}}{\partial {\bf h}}.
\end{align}

For every frozen neuron with $\sigma_{\mu} = 0$, there exists a left-eigenvector of the form $L_{\mu}^T = (0,\ldots,1,\ldots,0)$, nonzero only at location $\mu$,  that is a zero mode of the Jacobian, i.e. $L_{\mu}^{T} \mathcal{D} = 0$. 
For $N_{\rm f}$ frozen neurons, there will be $N_{\rm f}$ linearly independent zero left-eigenvectors. Since the dimension $D$ of the manifold is equal to the number of linearly independent zero modes, it will be precisely equal to the size of the frozen population, $D = N_{\rm f}$. We also note that bi-orthogonality implies that the right-eigenvectors $R_{\mu}$ corresponding to the zero eigenvalues are supported on exactly one frozen dimension, and the rest of the support is spread over the   active population.

 In the gRNN, the size of the frozen population is on average $N/2$, and thus the manifold is generically locally high-dimensional. In the ssRNN we can have a lower-dimensional manifold, not only because  the network can support manifolds for smaller $N$, but also due to the ability to achieve stable manifolds with a smaller fraction of frozen neurons by tuning $b$ in eq. \ref{eq:ssRNN-eom}. Thus, the ssRNN offers a way to robustly tune the dimensionality for a fixed $N$. 

\begin{figure}
\begin{centering}
\includegraphics[scale=0.58
]{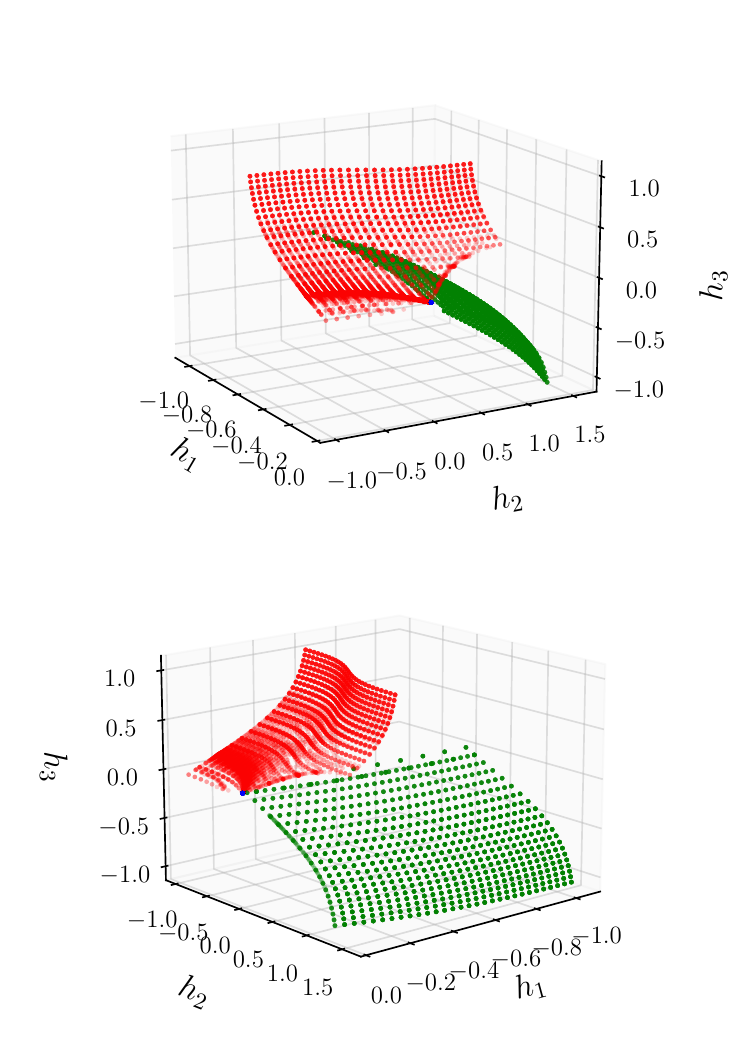}
\par\end{centering}
\caption{\label{fig:lowD-manifold} {\it Global geometry of the memory manifold} 
} Two views of an example $2-$D manifold generated by a gRNN with 3 units. The global structure consists of two ``sheets'' (red \& green) that are connected. Each sheet consists of a distinct subset of $2$ frozen units. The manifold was generated by sampling a grid of initial conditions in $[-1,1]^3$ and running the dynamics till a fixed-point was reached (further details about numerics in Appendix \ref{app-sub-globalgeometry}).
\end{figure}

In general, the global geometry of the manifold (embedded in $\mathbbm{R}^N$) can be quite complicated and even the dimensionality can change from one part of the phase space to the other. 
In Fig. \ref{fig:lowD-manifold}, we see an example of a ($2-$D) manifold embedded in $\mathbbm{R}^3$, for a gRNN with $N=3$. The manifold was generated by sampling initial conditions from a grid in $\mathbbm{R}^3$, and running the forward dynamics till convergence to a fixed-point (see Appendix \ref{app-sub-globalgeometry} for details of the numerics). We see that the global structure consists of two ``sheets'' (red \& green in Fig. \ref{fig:lowD-manifold}) that are connected -- each sheet corresponds to a different subset of (2) frozen units.   Note that the sheets themselves are continuous, and the discreteness is an artefact of the sampling.  Each sheet corresponds to a particular subset of units being frozen (i.e. a bit string $\pmb{\sigma}$). Which particular subset of units contribute to any particular sheet is controlled by the $W$ matrix, and the $J$ matrix controls the `stability of the sheet'.


The intuitive picture depicted in Fig. \ref{fig:lowD-manifold} hints at  the manifolds in higher dimensional gRNNs. The manifold will consists of ``sheets" labeled by a particular bit string $\pmb{\sigma}$ and locally homeomorphic to $\mathbbm{R}^{N_{\rm f}}$. The different sheets can intersect, and if they do, the intersection will be on the boundaries defined by the condition $W {\bf h}  = 0$. This condition defines a set of $N$ constraints which partition the space into $2^{N}$ distinct regions. And, each region corresponds to a bit string $\pmb{\sigma}$ labeling the frozen units. Will all the $2^{N}$ regions contain a sheet? Evidently no. The $W$ matrix partitions the phase space into regions labeled by a unique bit string $\pmb{\sigma}$ and each region will have a {\it zero} manifold -- i.e. a set of points satisfying the fixed-point criterion. However, the 
existence of a {\it sheet} depends on whether or not this zero manifold (or part of it) is stable/attractive, as dictated by the $J$ matrix. Thus, as mentioned earlier, the $W$ matrix controls the shape/geometry of the zero manifolds in the phase space, and the $J$ matrix controls their stability.  Only regions with stable sheets will contribute to the memory manifold. We elaborate on the condition for stability below.

\subsubsection{Metric, curvature and timescales on the manifold}
\label{subsubsec:metric-curvature-timescales}

 The discussion above paints a picture of the memory manifold as a union of attractive sheets. Here, we quantify intrinsic and extrinsic geometric properties of the individual sheets by means of the metric and curvature tensors, respectively. 
 The metric allows us to see how inputs move the activity along the manifold, and surprisingly, reveals relaxational dynamics with a wide range of timescales on a single sheet. 
   An input aligned with a right-zero-mode $I {\bf r}^{(\mu)}$  of the Jacobian will have support on only a single frozen coordinate (and on the entire active population), producing a change in only a single frozen direction $d h_{\mu}^{\rm f} = I dt$
which moves the network a distance $ds = \sqrt{g_{\mu\mu}} d h_{\mu}^{\rm f}$ along the manifold, where $g_{\mu \nu}$ is the metric tensor. More generally, an input spanning the whole frozen subspace, will produce a change in the state $ds^2 = |d\mathbf{h}|^2 = g_{\mu \nu} dh^{\rm f}_{\mu}dh^{\rm f}_{\nu}$. 

In Appendix \ref{app-geometry}, we show that the  metric tensor is given by 
\begin{align}\label{eq:metric-2}
    g_{\mu \nu} =  \delta_{\mu \nu} + \phi'(h^{\rm f}_{\mu})\phi'(h^{\rm f}_{\nu})\Gamma_{\mu \nu},
\end{align}
where $\Gamma = \left(J^{\rm f}\right)^{T} \Sigma^{-1} J^{\rm f}$, and $\Sigma = \mathcal{D}_{\rm a} \mathcal{D}_{\rm a}^{T}$ involves the ``active" Jacobian $\mathcal{D}_{a}$ defined in Eq. (\ref{eq:vRNN-jac}). We immediately note the explicit dependence of the metric on the non-linearity $\phi(\cdot)$. In particular, if $\phi(\cdot)$ is a piece-wise linear function, such as $\phi(x) = \max(0,x)$, then the metric is {\it locally flat}, i.e. proportional to the identity under a linear coordinate transformation. 

For the metric to be well-behaved, we need $\Sigma$ to be invertible, which in turn requires $\mathcal{D}_{\rm a}$ to be invertible. Failure of invertibility here indicates that the sheet will no longer be attractive. This will occur when the slope $ds/dh^{\rm f}_{\mu}$ becomes infinite along even a single direction
$h^{\rm f}_{\mu}$, thus yielding  a $(N_{\rm f} - 1)$-dimensional manifold boundary.
The invertibility of $\mathcal{D}_{\rm a}$ can also be studied using the local time constant $\tau({\bf h}^{\rm f})$, given by the norm of the inverse of the maximal real part of the spectrum of $\mathcal{D}_{\rm a}$,

\begin{align} \label{eq:vRNN-spectral-abcissa-nonMFT}
\tau({\bf h}^{\rm f}) =  \left(1-\frac{\mu_{\sigma} }{N_{\rm a}} \sum_{i }  \left( \phi'(h_{i}^{\rm a}({\bf h}^{\rm f})) \right) ^{2}  \right)^{-1}.
\end{align}

A generic point on a sheet will have a $\tau({\bf h}^{\rm f})$ determined from the MFT prediction in eq. (\ref{eq:vRNN-spectral-abcissa}). However, it is important to note that 
$\tau$ is not constant throughout the phase space, even on a single sheet; it takes its smallest value deep in the bulk and diverges $\tau ({\bf f}) \to \infty$ precisely on the boundary of a given sheet \footnote{ Deep in the bulk of a manifold, as ${\bf h }^f \to \infty$, $\phi({\bf h}^f)$ saturates and a mean-field theory gives $\Delta^{a} = \mu_{\sigma} C_{\phi}(\Delta^{a}) + 1 - \mu_{\sigma}$. This gives a bulk $\tau_{bulk}$, which is the largest value it can take on the manifold. Statistically, a given bit-string partition will not contain any stable attracting set if $\tau_{bulk}(\pmb{\sigma}) < 0$.}.
Importantly, $\tau({\bf h}^{\rm f})$ reflects the relaxation time of small perturbations off the sheet. Therefore, even on a single sheet $\mathcal{M}(\pmb{\sigma})$, there is a wide distribution of local relaxation times $\tau \in (\tau_{bulk}, \infty)$.

We can see from the metric (eq. \ref{eq:metric-2}) that the manifold is intrinsically mostly flat. However, it is interesting to know how it is embedded in the full ambient space of $\mathbbm{R}^{N}$. To access this, we look to the extrinsic curvature, 
which describes how rapidly the normal vectors to the manifold change along the manifold. For instance, if the normal vector remains constant, then the manifold is embedded as a hyperplane. If the normal evolves rapidly along the manifold, then it will appear highly curved. We show in Appendix \ref{app-geometry} that the extrinsic curvature is a fluctuating quantity (in the disordered connectivity), with zero mean and a variance which scales as $N^{-1}$. Since the eigenvalues of the extrinsic curvature tell us about the principal radii of curvature, we see that high-dimensional memory manifolds tend to have very low curvature. We provide the precise equation for this subleading variance in Appendix \ref{app-geometry}.
We point out also that the extrinsic curvature is proportional to the second derivative of the activation $\phi''(x)$, and is therefore exactly zero for piecewise linear activation functions.  The approximate flatness of the memory manifold implies that they can be effectively used to store multiple continuous memories without interference via integration. We substantiate this claim in the next section.

\section{A general purpose integrator}
\label{sec:integrator}

\begin{figure}
\begin{centering}
\includegraphics[scale=0.7
]{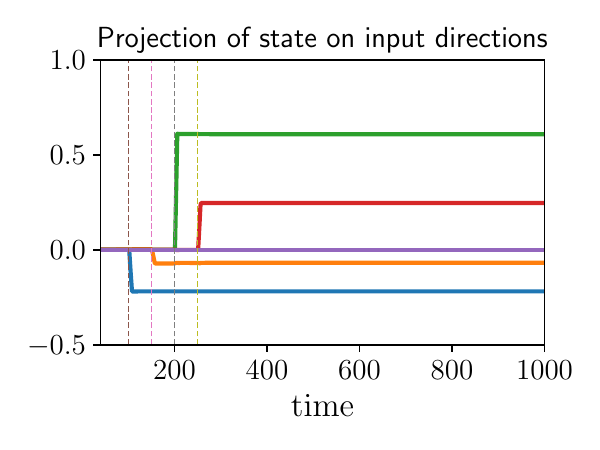}
\par\end{centering}
\caption{\label{fig:integrator} {\it Memory manifold can be used as an integrator:} Input pulses lasting 5 time units, with random amplitudes, are applied along randomly selected manifold directions starting at the times indicated by the vertical dashed lines. The (excess) projection of the state on the input directions shows that the input trace is integrated and retained in the network state $\mathbf{h}(t)$ for a long time. Moreover, the inputs along different directions do not interfere with each other. The network used is a ssRNN with $g=1.0, N=100, \gamma = 1.0, G=1.0$.
}
\end{figure}

One of the crucial functions a memory manifold can implement is that of a general purpose integrator: external input along the manifold directions will move the state on the manifold, and once the input ceases, the final state will store the integrated value of the input; however, traces of inputs that are not aligned to the manifold will decay away rapidly. Integrator function of this type forms the basis of many biologically important computations \cite{gold2007neural,brunton2013rats,burak2009accurate,tolman1948cognitive,nieh2021geometry}. 
It is well known that implementing an integrator in biologically realistic models requires fine-tuning and special symmetries \cite{goldman2010neural, seung1996brain,chaudhuri2016computational,bialek2012biophysics}. Frozen stabilisation might offer a possible way for biological systems to implement the integrator function over a wide range of parameters and interaction patterns. In Fig. \ref{fig:integrator} we see an example of this integrator function: inputs lasting a short duration ($5$ time units) are applied along different directions of the memory manifold produced by a ssRNN, at times indicated by the vertical dashed lines. The projection of the network state $\mathbf{h}(t)$ on the input directions shows that the state integrates the input and maintains it in memory for a long time; moreover, the traces of inputs along the different directions do not interfere with each other. Thus, different manifold directions could in principle be used for separate tasks \footnote{ The gRNN also offers this integrator functionality, albeit typically with a larger system size. }. This confirms our analysis of the extrinsic geometry, showing that the memory manifolds are sufficiently flat to permit the integrator functionality. 

Another salient feature of frozen stabilisation is that it allows a truly continuous integrator to be robustly implemented in systems with low dimensionality (e.g. a ssRNN with $N\sim20$). This is in contrast with the classical models for continuous attractors  in neural systems (e.g. \cite{ben1995theory,burak2009accurate}) where the number of fixed-points, and thus the integrator resolution, scale with the system size, $N$. Recent experimental work in the nervous system of insects has provided evidence for integrators with continuous attractor dynamics for representing angular orientation with a small number ($\sim 20$) of neurons \cite{kim2017ring}.  If we use the classical models in their original form, it raises the problematic question of the resolution of the angular representation. Thus, memory manifolds generated by frozen stabilisation are a potential solution for achieving  high-resolution continuous attractors with a small number of neurons.


\section{Long timescales without fine-tuning}
\label{sec:long-timescales}

Now we consider a problem that is closely related to the problem of memory: generating long timescales. To function as a memory, a system might require one long timescale, whereas other behaviours/functions might require a whole spectrum of timescales. We consider both these scenarios below. Regarding one long timescale, the problem can be stated as generating  timescales in the collective dynamics of a system, which are much longer than the intrinsic timescale of its units. Typical examples of achieving this require a system to be close to a bifurcation point in the parameter space -- i.e. a fine-tuning of parameters is needed. In general, achieving long timescales without fine-tuning is  challenging \cite{chen2020searching,bialek2012biophysics, seung1996brain}. On the other hand, processing inputs with multiple timescales requires responses with a spectrum of timescales, and this too is generally challenging to achieve. We will see below that frozen stabilisation leads to a wide range of relaxation times in response to perturbations.

\subsection{Long intrinsic timescales}

From the discussion of frozen stabilisation above, we know that for a wide range of parameters (e.g. $1 < g \lessapprox 3.27$ for the gRNN) and binary $\pmb{\sigma}$, the dynamics will flow to marginally-stable fixed-points, which implies an infinitely long correlation/autonomous timescale.  Thus, these long autonomous timescales are a consequence of the network self-organising to a critical state and do not require fine-tuning of parameters.

\begin{figure}[h]
\begin{centering}
\includegraphics[scale=0.45
]{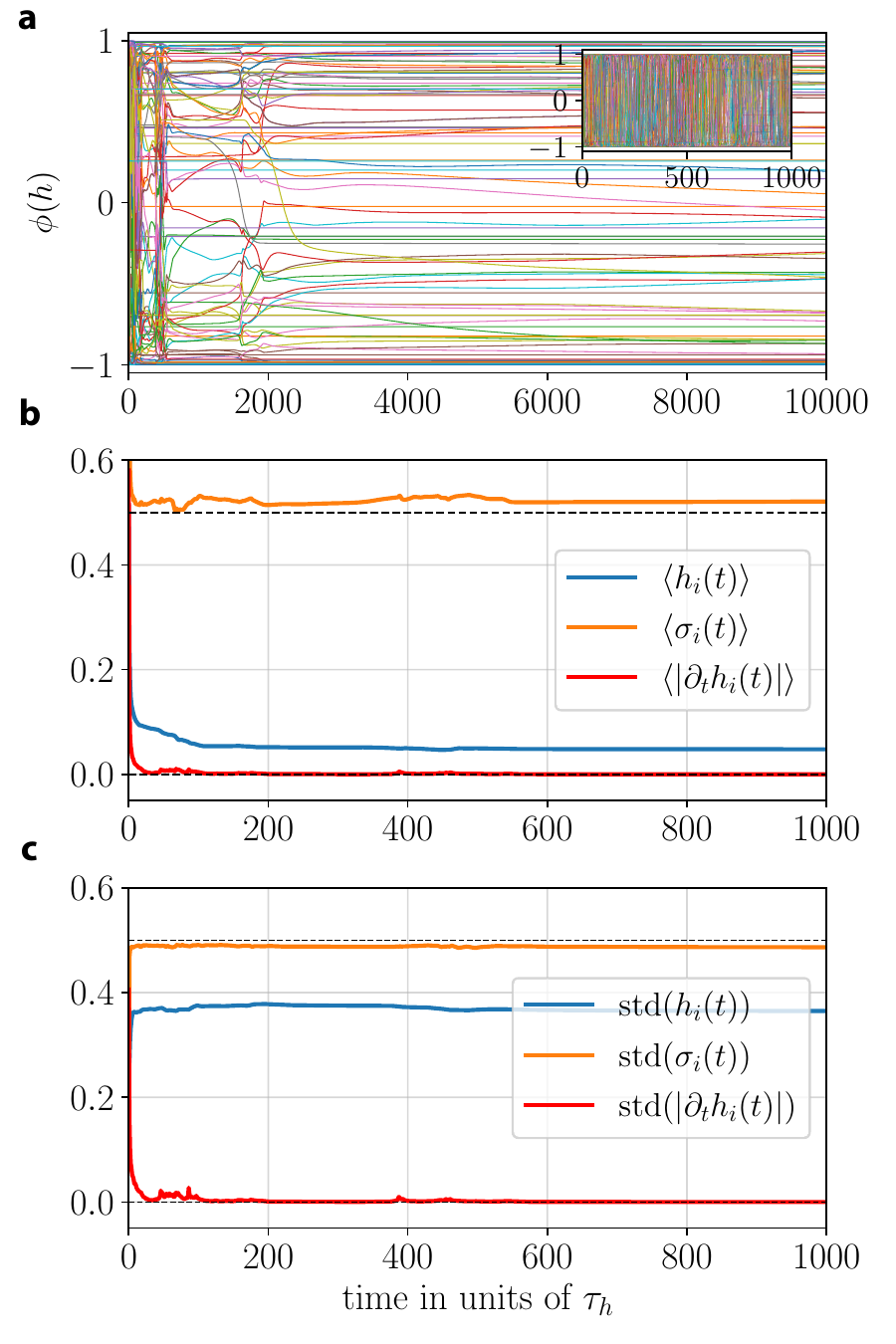}
\par\end{centering}
\caption{\label{fig:long-timescales} {\it Emergent long timescales:} a) Sample traces of a gRNN exhibiting long timescales in the collective dynamics : $N=1000,\alpha=50,g=2.0$; (inset) same network but with $g=4.0$ shows fast, chaotic dynamics. 
b,c) Mean(b) and standard deviation(c) of population averaged quantities, showing that the velocity ($\partial_t h_i(t)$) for all neurons are rapidly suppressed after short transients. Note that half of the $\sigma_i$ are open are open at any given instant.
}
\end{figure}

We now extend the discussion of self-organised generation of long timescales to the case when $\pmb{\sigma}$ is sensitive but not strictly binary. We can parametrise the sensitivity with a gain $\alpha$ such that $\sigma(x) = [1 + \exp(-\alpha x)]^{-1}$; $\alpha \to \infty$ would give the binary case. For concreteness, let us consider the case of the gRNN with large-but-finite $\alpha$.  In Fig. \ref{fig:long-timescales}a, we see that for large $\alpha$ the dynamics for the network still exhibit timescales that are many orders of magnitude larger than the intrinsic timescale of a neuron ($\tau_h$). Like the strictly marginally-stable state these long timescales are a self-organised phenomenon and not simply the result of having a sensitive $\pmb{\sigma}$; indeed for $g$ outside the range supporting marginal stability, the exact same network exhibits overall fast chaotic dynamics, with only intermittent short-lived freezing of activity (Fig. \ref{fig:long-timescales}a, inset). 

It is instructive to study features of this self-organised behaviour using the ``velocities'' $| \partial_t h_i(t) |$. In Fig. \ref{fig:long-timescales}b,c we see that the characteristic feature of this dynamical state is that after short transients the  velocities $\partial_t h_i(t)$ are quickly suppressed to very low values for {\it every} neuron (see red lines in Fig. \ref{fig:long-timescales}b,c), even though the population averaged values of other quantities such as $\langle h_i(t) \rangle$ may be biased away from zero (the expected value) \footnote{ The positive bias in this case is a consequence of initial conditions. The persistence of a finite fraction of the initial bias in the mean of $h$ is an indication of the emerging slow manifold.}.  We also see that roughly half the $\pmb{\sigma}_i$ are open and roughly half are closed -- compare the mean and standard-deviation of $\{ \sigma_i(t) \}$ (orange lines in \ref{fig:long-timescales}b,c). This means that half the $h-$population is free to evolve. Thus, the collective dynamics gives us access to long timescales over a wide range of values for $g$ and $\alpha$. As a final point, how long are the long timescales in this regime of large-but-finite $\alpha$? A simple diffusion argument suggests that these timescales should be closely related to the timescale for the $\sigma_i$ switching which will scale as $\tau_{switch} \sim \alpha^2 \langle \phi(h)^2 \rangle$.


\subsection{Wide range of relaxation times}\label{sec:SOC}

\begin{figure}
\begin{centering}
\includegraphics[scale=0.58
]{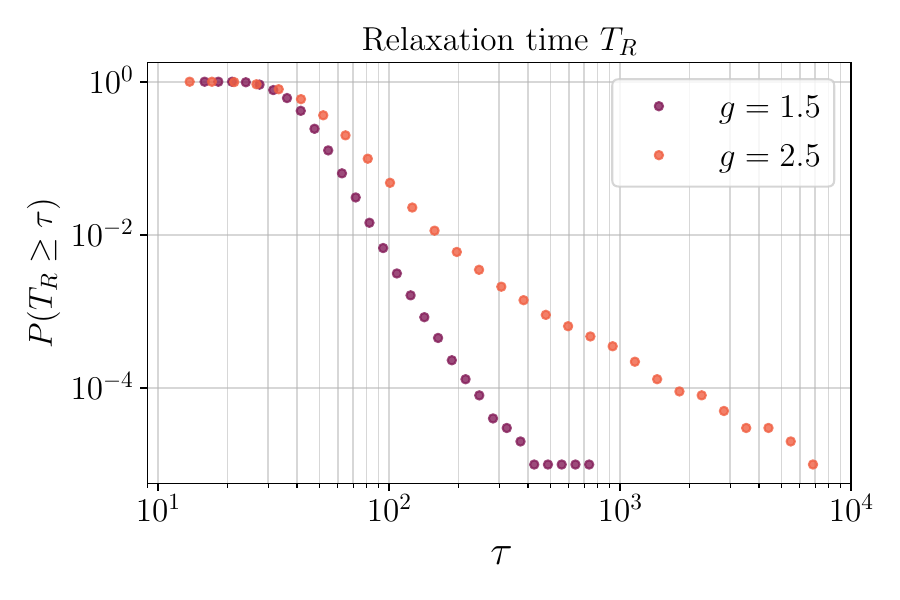}
\par\end{centering}
\caption{\label{fig:relaxation-times} {\it Wide range of relaxation times:} Complementary cumulative distribution functions of the relaxation time $T_R$ to perturbations for two different values of $g$ show a lower cut-off and a heavy tail, implying a wide range of timescales. The numerics used a gRNN with $N=100, \alpha = \infty$. The perturbations were random Gaussian vectors with a variance of $0.25$, and the relaxation time was calculated as time required for the time-derivatives to reach 0 (within solver tolerance). 
}
\end{figure}

Having considered  timescales in the autonomous dynamics above, we now study the timescales present in the {\it responses} of a  system exhibiting frozen stabilisation to random perturbations. Specifically, we look at the timescales with which the marginally stable system relaxes back to a fixed-point when perturbed in a random direction. In Fig. \ref{fig:relaxation-times}, we see that the relaxational timescale ($T_R$) exhibits heavy-tailed distributions. These distributions bear similarity to those of  ``avalanche sizes'' commonly observed in self-organised critical phenomena \cite{jensen1998self}. However, there are some notable differences. Firstly, unlike the scale-free behaviour characteristic of  critical avalanches, the relaxation here cannot be truly scale-free. Indeed, the very mechanism of frozen stabilisation implies there should be a timescale present in the problem that is given by the spectral abscissa of the ``stabilised" part of the spectra (circled in green in Fig. \ref{fig:Jac_schematic}). We expect this to dominate the short-time behaviour. Secondly, we do not presently have an analytical handle on the functional form of the tail behaviour. Thus, it is possible that we do not observe power-law tail behaviour typical of critical avalanches -- all we can demonstrate is a wide range of relaxation times.

What is the mechanism that gives rise to the wide range of relaxational timescales? There are two aspects of the geometry that contribute to this. Firstly, as mentioned above, the timescale of relaxation to a small perturbation off the sheet is governed by the local time constant determined from the active Jacobian Eq.(\ref{eq:vRNN-spectral-abcissa-nonMFT}). 
This timescale varies over the sheet and takes it smallest value in the bulk of the sheet, and diverges at the boundary of a sheet. So, even within a single sheet there is a wide range of relaxation timescales to small perturbations. The second aspect of the geometry contributing to the wide range of timescales is ``jumping'' between sheets after a moderate perturbation. To see this, we note that in a frozen stabilised network  there is a threshold function which spontaneously segregates the population into frozen and active.
This threshold function thus takes the activity of a network and returns a $N$-bit string $\pmb{\sigma}$, with $\pmb{\sigma}_{i} \in \{0 , 1 \}$ indicating the participation of each neuron. A moderately large perturbation will be able to flip some subset of bits, pushing the network away from the sheet. Subsequently, in relaxing back to the manifold -- possibly to another sheet -- more bits will need to be flipped, causing a branching process. In this way, the relaxation has semblance to critical avalanches, which exhibit a wide range of timescales. 



\section{Discussion}
\label{sec:discussion}

In this work we address two related challenges that arise in the context of neural networks, and dynamical models of biological systems more broadly: i) how to store a memory trace of continuously varying quantities; and ii) how to generate a spectrum of long timescales in the collective dynamics.
The ability to implement memory and generate long timescales underlies several  functions critical for survival such as evidence accumulation from a noisy signal \cite{gold2007neural,brunton2013rats}, path integration based on velocity cues \cite{burak2009accurate, seung1996brain}, forming cognitive maps \cite{tolman1948cognitive,nieh2021geometry}, just to name a few.
Prior approaches to achieving this have typically relied either on systems being carefully poised at the boundary between stability and instability, thus requiring tuning parameters to a special value, or on special symmetries in the connectivity. 
Marginal stability without symmetries or parameter fine-tuning requires the system to dynamically self-organise into a critical state. This is a highly non-trivial phenomenon, and apart from specific models , more general principles are hard to come by.

Here we present a principle that we refer to as {\it frozen stabilisation}, which allows a family of models to self-organise to a critical state exhibiting a continuous memory manifold and a wide spectrum of response timescales. The principle is robustly implemented over a range of parameters and doesn't require any special symmetries in the interactions/connectivity. 
Previous work has managed to attain continuous attractors with seemingly unstructured interaction matrices \cite{darshan2021learning}. However, these arise as a result of learning and are sensitive to perturbations in the learned weights -- i.e. they are fine-tuned. 

A salient feature of frozen stabilisation is that it can be implemented in small systems yielding  a true continuum of fixed-points, which is in contrast to prior models where the manifold is approximated by a discrete set of fixed points that scale with the system size (see \cite{noorman2022accurate} for another recent model which implements continuous attractors in small networks, albeit with some fine-tuning). 
This is relevant to debates about implementing a continuous memory with a small number of neurons, which have become important in light of the experimental results on the fly navigation system \cite{kim2017ring}. To our knowledge, frozen stabilisation is the first example of a principle that allows robust implementation of continuous attractors in small systems. 

The memory networks exhibiting frozen stabilisation can be used as general purpose integrators, where inputs along the different manifold directions are independently integrated and retained in memory without interfering with each other. 
 Indeed,  in machine learning tasks requiring evidence accumulation, trained gated RNNs solve the tasks by constructing continuous attractor manifolds, and it is likely that frozen stabilisation might be at play in facilitating these continuous attractors 
  \cite{maheswaranathan2019reverse,maheswaranathan2020recurrent,maheswaranathan2019universality}. More generally, it is possible to leverage the memory networks produced by frozen stabilisation to provide a ``memory reservoir'' for other RNNs in tasks requiring information processing over long timescales \cite{kim2023trainability}.

 The geometry of the memory manifolds is dictated by the form of the freezing/slowing function that spontaneously splits the full network into active and frozen subpopulations. For a random freezing function, we get a union of typically disjoint ``sheets'', where each sheet functions as an attractive memory manifold subserved by a random subpopulation of frozen neurons. It is interesting to note that experiments on spatial maps in the rodent hippocampal formation also reveal the presence of multiple maps, possibly subserved by random neuronal populations \cite{wills2005attractor, derdikman2011manifold,alme2014place,sheintuch2020multiple}. Models of continuous attractors with multiple maps or ``charts" were proposed by \cite{samsonovich1997path}, and and studied further for independent \cite{battaglia1998attractor,monasson2013crosstalk} and correlated \cite{romani2010continuous} maps. The storage capacity is greatly diminished for multi-map attractor networks, whereas our memory manifold consists of {\it exponentially many} (in $N$) stable sheets. Moreover, each sheet in FS exhibits a wide range of relaxational timescales for responses. The presence of multiple sheets and a wide range of response timescales are an emergent property of FS and do not depend on  precise details of the couplings or the model implementation.  In follow-up work, we have shown that it is possible to tune the shape of the memory manifold by learning an appropriate freezing function (e.g. learning the $W$ matrix in the gRNN) \cite{kim2023trainability}. We leave exploring this direction further for future work.

We discussed the principle of FS using examples of specific neural network models, and a salient feature of these models was the binary $\pmb{\sigma}$ variable that  spontaneously freezes a part of the system depending on the instantaneous system state. One natural question that arises is whether frozen stabilisation can be implemented without the binary $\pmb{\sigma}$?  If we consider a more general class of dynamical systems specified by $\partial_t \mathbf{h}(t) = F[ G(\mathbf{h}) ]$, where $F$ is a function acts element-wise, and has some regions of its domain where $F(x)=0$, then units for which $G(h_i)$ fall in those regions will be momentarily frozen, and we can express the dynamics {\it equivalently} using gates as discussed in Sec. \ref{sec:frozen-stab-principle}. For instance, when  $F= \textrm{ReLU}(x) \equiv \max(0,x)$, we can express the dynamics using gates since ${\rm ReLU}(x) = \sigma(x) x$,  where $\sigma(x)$ is the gate/step function. The key requirement seems to be the ``threshold'' behaviour of $F$ which makes it inactive over certain regions of its domain. This can, potentially, be implemented in neurons with hysteresis \cite{koulakov2002model,goldman2003robust}. A perfectly step-like gate is also not strictly necessary. The really crucial ingredient which step-like gates readily provide is a clear separation of timescales which is produced dynamically via state-dependent time constants. 
 
Such state-dependent functional characteristics, which are a crucial feature of our theory, are in fact widely observed in real neurons \cite{koch1984biophysics,koch2004biophysics, silver2010neuronal}, and can be linked to many biophysical mechanisms including nonlinear dendritic integration and neuromodulation. Similarly, \cite{gao2020neuronal} found a wide range of timescales in cortex that were not static, but ``functionally dynamic", in that they were found to selectively increase in response to working memory tasks. \cite{marom2010neural} has even argued that a notion of a neuronal time constant should be replaced with a picture of adaptive timescales in neurons. Ref. \cite{gutig2009time} went further and showed that adaptive time constants in conductance neurons can make the response of the neuron invariant to time-warped inputs, illustrating a compelling computational advantage of such a feature. Finally, we demonstrate a mechanism in Appendix \ref{app-balance} which would give rise to an effective rate network description with a dynamical timescale similar our model Eq. \ref{eq:large-gRNN}.  Following the proposal of negative-derivative feedback in \cite{lim2013balanced}, we show that {\it nonlinear} neurons which receive excitatory and inhibitory feedback, balanced in strength but differing in intrinsic timescales, can experience an effective state-dependent time constant.
 
 More generally, the discussion of frozen stabilisation through the lens of the Jacobian spectrum suggests that a larger family of dynamical systems could yield the form of the spectrum required to exhibit frozen stabilisation. We leave characterising  the full class of systems that can implement the FS principle  for future work. 

One natural concern is that since each neuron can, in principle, achieve a slow time-constant, each individual neuron can act as an integrator. From this perspective, it 
may seem like the emergence of the continuous attractor is less of network phenomenon than a property of single neurons.  However, to alleviate this concern, as we point out in Appendix \ref{app:notch-gating}, the network effects of frozen stabilisation can be absolutely crucial in establishing functionally relevant integrator behaviour. In particular, we show that even when each neuron is a very poor integrator, wiring them up in a network will produce superior integrators via the emergence of high dimensional manifolds. Moreover, single neuron timescales don't explain the large number of manifolds and the wide spectrum of relaxation timescales -- both emergent properties of frozen stabilisation.

In summary, we have presented a principle that allows for robust implementation of continuous memories and a wide range of response timescales in a family of neural networks, without fine-tuning parameters or using special symmetries. 
The continuous memory manifolds produced by this principle exhibit a true continuum of fixed points and can be implemented in small networks. 
Such memory networks could be useful to model biological implementations of integrators or cognitive maps. These networks could also serve as ``memory reservoir'' modules for use in larger machine learning architectures requiring information processing over long timescales.


\section{Materials and methods}
\label{sec:methods}

\noindent{\bf Dynamical mean-field theory (DMFT):}
Appendix \ref{app-ssRNN} presents details of the DMFT used to predict the parameter values quoted in Fig. \ref{fig:vRNN_biases} which give rise to marginal stability, as well as to derive the static to dynamic noise ratios \ref{eq:noise_ratio_grnn} and \ref{eq:noise_ratio_ssrnn}.

\smallskip
\noindent{\bf Manifold geometry:}
Appendix \ref{app-geometry} gives details of the intrinsic and extrinsic geometry of the memory manifolds, providing a derivation of Eqs.  \ref{eq:metric-2} and \ref{eq:vRNN-spectral-abcissa-nonMFT}, as well as proving that the mean extrinsic curvature vanishes as $N^{-1}$. 

\smallskip
\noindent{\bf Notched gating and network effects:}
Appendix \ref{app:notch-gating} presents simulations and studies integrator functionality of a gRNN with a ``notched'' gating function
\begin{align} \label{eq:notch-gRNN-methods}
    \sigma_{notch}(x) =& 
    \left\{
\begin{array}{ll}
      0 & 0 \leq |x| \leq \beta/2 \\
      1 & \textrm{o/w}\\
\end{array} 
\right. ,
\end{align}
for $\beta$ smaller than the input interval, the individual neuron is a poor integrator. Nevertheless, the network is able to integrate longer input intervals due to the emergence of robust memory manifolds via frozen stabilisation. 



\smallskip
\noindent{\bf Negative-derivative feedback and gating:}
In Appendix \ref{app-balance}, we extend the analysis of \cite{lim2013balanced} and show how a network in which excitation and inhibition are balanced in strength, but offset in their timescale $\Delta \tau = \tau_{E} - \tau_{I} > 0$, one can recover an effective rate equation 
\begin{align*}
    \tau \dot{r} & \approx  - r - \Big( J \Delta \tau  \phi'(r) \Big) \dot{r} + I,
\end{align*}

in which the damping coefficient is only constant if the activation function $\phi(\cdot)$ is linear. This can be rewritten as a state-dependent gating by defining $\sigma(r) = \left( \tau + J \Delta \tau \phi'(r)\right)^{-1}$.


\smallskip
\noindent {\bf Network Simulations:}
For all network simulations, unless otherwise explicitly stated, we used randomly generated matrices $W$ and $J$, with zero mean and variance $1/N$. All figures showing the results of simulations indicate the parameter used. For sufficiently large systems, we expect self-averaging to occur, in the sense that the specific realizations of $J$ and $W$ are unimportant for the qualitative (and even quantitative) results. 

For Fig. \ref{fig:integrator}, we allowed the network with random initial condition to settle to a fixed point, around which we evaluated zero modes of the Jacobian matrix $\mathcal{D}[{\bf h}]$, which depends on the state vector ${\bf h}$. The zero modes we used were the right-eigenvectors of the Jacobian that satisfy $\mathcal{D} {\bf R}_{\mu} = 0$, where $\mu = 1, ..., N_{\rm f}$. We chose five of these zero modes arbitrarily, and project input as brief pulses along these directions ${\bf I} = I(t) {\bf R}_{\mu}$.

\vspace{20pt}

\begin{acknowledgments}

\noindent {\bf Funding:} KK is supported by a C.V. Starr Fellowship and a CPBF Fellowship  (through NSF PHY-1734030). During the completion of this work, TC was supported by the Eric and Wendy Schmidt Membership in Biology, the Simons Foundation (891851, TC), the Starr Foundation Member Fund in Biology at the Institute for Advanced Study, and the Initiative for Theoretical Science at the Graduate Center, CUNY. 

\bigskip

\noindent {\bf Author Contributions:} All authors contributed equally to all aspects of the manuscript. 

\bigskip
\noindent {\bf Competing Interests} 
All other authors declare they have no competing interests.

\bigskip
\noindent {\bf Data and materials availability:} Code to reproduce all the figures can be provided upon request.

\end{acknowledgments}


\nocite{*}

\providecommand{\noopsort}[1]{}\providecommand{\singleletter}[1]{#1}%
%


\newpage
\appendix

\begin{widetext} 

\begin{center} { \Large Supplementary Materials for}\\
\smallskip
{\bf \large Emergence of Memory Manifolds}\\
\smallskip
{  Tankut Can and Kamesh Krishnamurthy}
\end{center}

\bigskip 
\section{Mean-field theory and dynamics of model networks} \label{app-ssRNN}

\subsection{Details of the gated RNN (gRNN)}

In this section we provide details of the marginally stable regime in the gRNN discussed in Sec. \ref{sec:frozen-stab-gRNN}. Stability of the fixed-points are determined by examining the spectrum of the instantaneous Jacobian $\mathcal{D}$, which describes the linearised dynamics. If the spectral abscissa, or maximal value of the real part of the Jacobian spectrum (see Fig.(\ref{fig:Jac_schematic})), is less/greater than zero, then the fixed point is stable/unstable.The Jacobian for the gRNN is given by
\begin{align}
    \mathcal{D} \equiv \frac{\partial \dot{ \bf h}}{\partial {\bf h}} = [ \pmb{\sigma}] \left( - \mathbbm{1} + J [ \pmb{\phi}']\right),
\end{align}
where $[\mathbf{x}]$ denotes a diagonal matrix with the vector $\mathbf{x}$ on the diagonal.  There are no stable fixed-points for $g > 1$ in the absence of gating (i.e. for $\sigma = 1$) . However, a binary $\sigma$ leads to many  {\it marginally stable} fixed points, over a range of values of $g$, with the spectral abscissa of the Jacobian exactly equal to zero. The condition for this marginal stability relies on the expected value of the gate variable $\mu_{\sigma} = \langle \sigma \rangle$, and is given by 
\begin{align} \label{eq:large-gRNN-MS}
   \int Dx \left( \phi^{\prime}\left( 
   \sqrt{\Delta_h}x
   \right) \right)^2  < &  \: 1/\mu_{\sigma},
\end{align}
where $Dx = \exp( -x^{2}/2)/\sqrt{2\pi}$ is the standard Gaussian measure, and  $\Delta_h$ is the mean-field (fixed-point) variance obtained from the self-consistency condition : $ \Delta_h = \int D x \phi(\sqrt{\Delta_h}x)^2$ \cite{krishnamurthy2020theory}. This corresponds to a range  $1<g \lessapprox 3.27$,  over which the network will self-organise to a marginally stable state, and the dynamics will flow to one of the many marginally-stable fixed points. Furthermore, in this regime, there occurs an extensive number $\approx N (1 - \mu_{\sigma})$ of slow modes which become exactly zero modes in the binary gate limit. Note that the matrices $(J, W)$, are completely unstructured.

We now calculate the typical ($\sigma_{\beta}^{DMFT}$) and critical ($\sigma_{\beta}^{*}$) variance of the static noise/biases in the gRNN as described in Sec. \ref{subsec:frozen-stab-mech-gRNN}. Recall, that the dynamics of the network in the marginally stable regime are described as in eq. \ref{eq:two-pop-eom}. We now proceed to calculate two quantities: i) the critical values of the variance of external biases that will stabilise the active population and ii) the typical value of the variance of biases provided by the frozen population in the gRNN. The critical value of the variance of biases, $\sigma_{\beta}^{*}$, that will stabilise the dynamics of the active population is determined by the stability of the Jacobian for the active system at the fixed-point, which is given by
\begin{align} \label{eq:vRNN-jac}
    \mathcal{D}_{\rm a} = -\mathbbm{1} +   J^{\rm a} [\phi^{\prime}({\bf h}^{\rm a})].
\end{align}
For large $N$, the spectrum is self-averaging, and using mean-field theory (MFT), we find that the maximal real part of the spectrum, or the spectral abscissa $\alpha(\mathcal{D}_{\rm a})$, of the active Jacobian is given by 

\begin{align} \label{eq:vRNN-spectral-abcissa}
\alpha(\mathcal{D}_{\rm a}) =  \mu_{\sigma} 
\underbrace{\int Dx \left( \phi'(\sqrt{\Delta^{\rm a}} x)\right)^{2}}_{C_{\phi'}(\Delta^{\rm a})}
- 1,
\end{align}
 where the variance of the active population $\Delta^{a}$  is determined self-consistently using the MFT as the solution to
\begin{align} \label{eq:vRNN-MFT}
\Delta^{\rm a} = \mu_{\sigma} C_{\phi}(\Delta^{\rm a}) + \sigma_{\beta}^{2}.
\end{align}
In order to get the critical variance, $\sigma_{\beta}^{*}$, we solve for $\alpha(\mathcal{D}_{\rm a}) = 0$ using  eqns. \ref{eq:vRNN-spectral-abcissa},  \ref{eq:vRNN-MFT}. This value will be a function of $g$ and $\mu_{\sigma}$; however, we set $\mu_{\sigma} = 1/2$ in Sec. \ref{subsec:frozen-stab-mech-gRNN}. The local time constant of Eq. (\ref{eq:vRNN-spectral-abcissa-nonMFT}) is given by $\tau = -1/\alpha$. 

To calculate the {\it typical} value of the variance of biases provided by the frozen population in the gRNN, we assume that the distribution of frozen neurons is the same as that of the active neurons. This is a kind of statistical permutation invariance.  In this case, the MFT implies that
\begin{align} \label{eq:gRNN-bias-dmft}
\sigma_{\beta}^{DMFT} = \frac{C_{\phi}(\Delta)}{2},
\end{align}
 where $\Delta$ is the solution to $\Delta = C_{\phi}(\Delta)$.

\subsection{Analysis of self-stabilised RNN}

In this section, we calculate the stability of the ssRNN introduced in Eq. (\ref{eq:ssRNN-eom}). The phase space of the ssRNN is partitioned into ``cells" labeled by one of $2^{N}$ bit strings $\pmb{\sigma}$, each containing a separate ``map" or ``sheet" of the memory manifold. Crossing a boundary between two cells results in a bit-flip. The step-wise change in $\sigma$ will cause the Jacobian to be singular at the boundaries. To avoid this complication, our analysis will focus on regions away from the boundaries which have a well-defined bit-string. This should provide typical behaviour in the bulk of a given sheet. From this we can determine what fraction of sheets are stable. Away from the cell boundaries, the ssRNN instantaneous Jacobian is

\begin{align}
    \mathcal{D}_{ij} = \partial \dot{h}_{i}/\partial h_{j} = \sigma_{i} \left( - \delta_{ij} + J_{ij} (\phi'(h_{j}) r_{j}(t) \right).
\end{align}
Since $\partial_{{\bf h}} \pmb{\sigma} = 0$ away from the  cell boundaries, we have also that $\partial_{{\bf h}} {\bf r} = 0$ in these regions. Due to the freezing function $\sigma_{i}$, the Jacobian will assume a block structure. Without loss of generality, we assume that $\sigma_{i} = 1$ for $i = 1, ..., N_{\rm a}$, and zero otherwise. In this case, 

\begin{align}
    \mathcal{D} = \left(\begin{array}{cc}
    - 1 + J^{\rm a} [ \pmb{\phi}'\odot \pmb{r}] & -1 + J^{\rm f} [ \pmb{\phi}' \odot \pmb{r}]\\
    0 & 0 \end{array}\right),
\end{align}
where $\odot$ is the Hadamard product denoting element-wise multiplication of two vectors. The eigenvalue spectrum will then have $N_{\rm f} = N - N_{\rm a}$ zero eigenvalues. The rest of the eigenvalues are given by the upper-left block. For large $N_{\rm a}$, the distribution of these eigenvalues will follow a circle law with a support circumscribed by the curve $\lambda \in \mathbbm{C}$ satisfying

\begin{align}
    |\lambda + 1|^{2} =  \mu_{\sigma} \langle \phi'^{2} r^{2}\rangle_{\rm a}\label{eq:ssRNN-spectral-curve-2},
\end{align}
where $\mu_{\sigma} = N_{\rm a}/N$ is the instantaneous fraction of active neurons, and for short-hand we have denoted the empirical mean by an expectation value
\begin{align}
    \langle \phi'^2 r^{2} \rangle_{\rm a} \equiv \frac{1}{N_{\rm a}} \sum_{i = 1}^{N_{\rm a}} \phi'^{2}(h_{i}) r_{i}^{2}.
\end{align}

The expression (\ref{eq:ssRNN-spectral-curve-2}) follows from a straightforward application of techniques from non-Hermitian random matrix theory (we follow e.g. \cite{feinberg1997non,chalker1998eigenvector}). Since the expectation is evaluated over the active coordinates, $r_{i} = \gamma (1 - \mu_{\sigma})$. By then requiring the spectral abscissa of these bulk eigenvalues to be strictly negative implies the following criterion for marginal stability

\begin{align}
   \gamma^{2} \mu_{\sigma} (1 - \mu_{\sigma})^{2} \langle \phi'^{2} \rangle_{\rm a}\le 1 .
\end{align}

{\magenta


}

In order to estimate the phase diagram for the ssRNN, we utilize mean-field theory to evaluate the correlation function $\langle \phi'^{2}\rangle_{\rm a}$. Let us define the correlation function

\begin{align}
 \Phi(\Delta) = \int Dx \phi(\sqrt{\Delta} x)^{2},    
\end{align}
with $Dx = \exp(- x^2/2)/\sqrt{2\pi}$. Let us further define the empirical average of the frozen neurons. 

\begin{align}
    \Delta_{\rm f} = \frac{1}{N_{\rm f}} \sum_{i = N_{\rm a} + 1}^{N} (h_{i}^{\rm f})^{2}.
\end{align}

Then the mean-field theory gives us an implicit equation for the variance of the active population

\begin{align} \label{eq:ssRNN-MFT}
    \Delta_{\rm a} = & \: \gamma^{2}\mu_{\sigma}(1 - \mu_{\sigma})^2 \Phi(\Delta_{\rm a}) + \gamma^{2} G^{2} \mu_{\sigma}^{2} (1 - \mu_{\sigma}) \Phi(\Delta_{\rm f}) .
\end{align}

We look for solutions for which the frozen neurons follow the same distribution as the active neurons, i.e. $\Delta_{\rm a} = \Delta_{\rm f}$. This should give us the typical behaviour on a particular map, and allow us to build an approximate phase diagram. Under this condition, we have

\begin{align} \label{eq:ssRNN-MFT}
    \Delta_{\rm a} = & \: \gamma^{2}\mu_{\sigma}(1 - \mu_{\sigma}) ( 1+ (G-1) \mu_{\sigma}) \Phi(\Delta_{\rm a}) ,
\end{align}

which has nonzero $\Delta_{a}$ solutions only when the RHS has a slope at the origin greater than one. Applying this condition implies that nonzero solutions will exist for
\begin{align}
    \gamma^{2} \mu_{\sigma} (1 - \mu_{\sigma}) ( 1+ (G-1) \mu_{\sigma})  g^{2} \ge  1.
\end{align}

The curve set by the equality with $G = 1$ is plotted as solid blue in Fig. (\ref{fig:ssRNN-phase}). Marginally stable fixed points are only possible below the dashed blue curve, and these fixed points are generically nontrivial only above the solid blue line. 


We stress here that the phase diagram only makes sense for large $N$. However, the qualitative features appear to survive for small $N$. In particular, our phase diagram indicates that the ssRNN is better able to stabilise lower dimensional manifolds, which becomes more relevant for smaller $N$. Indeed, we find an ssRNN with ~50 neurons also shows robust frozen stabilisation (see Fig. \ref{fig:ssRNN-Jacobian}). 

We also quickly sketch the derivation of the noise ratio (\ref{eq:noise_ratio_ssrnn}). The freezing function will spontaneously divide the population in frozen $h^{\rm f}$ and active $h^{\rm a}$. Writing out the equations of motion in this basis gives

\begin{align}
\frac{d {\bf h}^{\rm a}}{dt} &= - {\bf h}^{\rm a} + \gamma (1 - \mu_{\sigma}) J^{\rm a} \phi({\bf h}^{\rm a}) + \gamma \mu_{\sigma} J^{\rm f} \phi({\bf h}^{\rm f}) ,
\end{align}
from which we may arrive at the effective mean-field stochastic ODE which describes the {\bf ssRNN  DMFT}
\begin{align}
&\dot{h}^{\rm a}  = - h^{\rm a} + \gamma (1 - \mu_{\sigma}) \sqrt{\mu_{\sigma}} \eta(t) + \gamma G \mu_{\sigma} \sqrt{1 - \mu_{\sigma}} \beta^{\rm f},  
\end{align}

Comparing this to the {\bf gRNN DMFT }
\begin{align}
 &\dot{h}^{\rm a} = - h^{\rm a} + \sqrt{\mu_{\sigma}}\eta(t) + \sqrt{1-\mu_{\sigma}}\beta^{\rm f}, 
\end{align}
allows us to see that the ratio $R = \sigma_{stat}^{2}/\sigma_{dyn}^{2}$ of static to dynamical noise between the two networks will be related

\begin{align}
    R_{ssRNN} = G^{2}\frac{\mu_{\sigma}^{2}}{(1 - \mu_{\sigma})^{2}} R_{gRNN}  = G^{2} R_{gRNN}^{-1}.
\end{align}
Recall that a larger ratio facilitates stabilisation. Therefore, the ssRNN works better at stabilisation precisely when the gRNN might struggle. This intuition is supported by Fig.(\ref{fig:ssRNN-phase}), which shows the range of gain parameters $g$ for which a manifold of dimension $N d = N ( 1- \mu_{\sigma})$ can be stabilised.


\begin{figure}
\begin{centering}
\includegraphics[scale=0.53
]{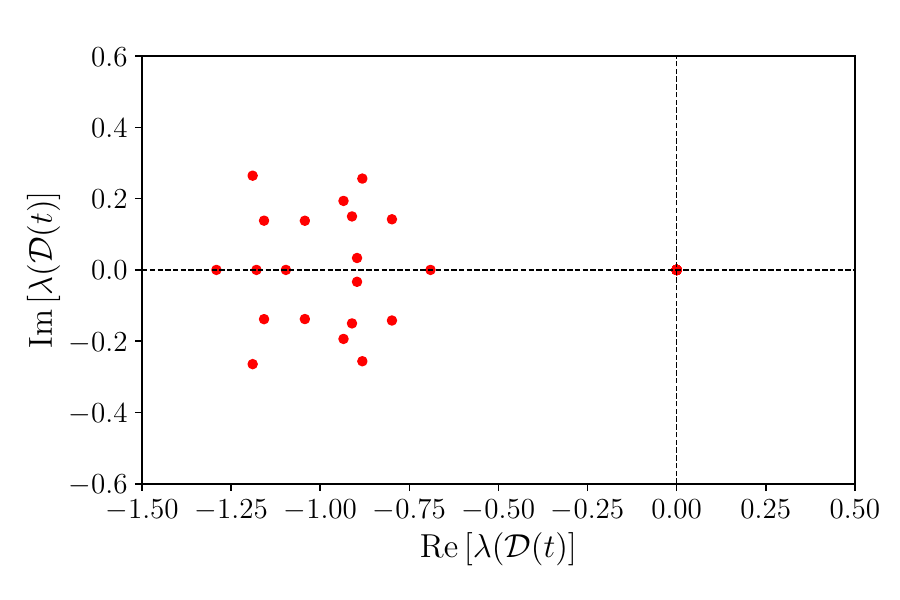}
\par\end{centering}
\caption{\label{fig:ssRNN-Jacobian}  {\it Instantaneous Jacobian spectrum for ssRNN with 50 neurons. Parameters used were: $N=50,g=1.0,\gamma = 1.0, G =1.0, b=0.0$}
}
\end{figure}

\begin{figure}
\begin{centering}
\includegraphics[scale=0.8
]{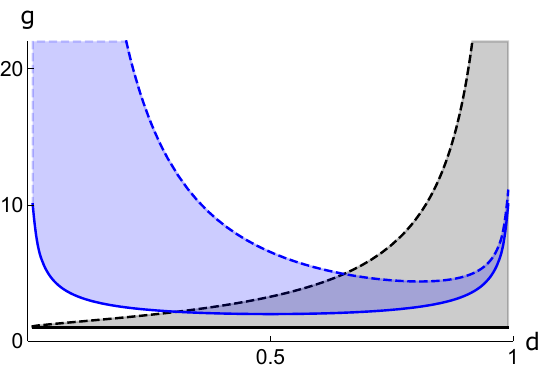}
\par\end{centering}
\caption{\label{fig:ssRNN-phase} The memory manifold is comprised of a set of sheets of different dimensions. For fixed gain parameter $g$, we determine all the dimensions $N d = N_{\rm f}$ which are stable in the mean-field. The blue region corresponds to ssRNN, showing that for increasing $g$, the memory manifold will be dominated by smaller $d$ sheets. The gray region is for the gRNN, which shows the opposite behaviour, where the memory manifold is overwhelmed by higher $d$ sheets for increasing $g$.  
}
\end{figure}


\section{Geometry of the memory manifold} \label{app-geometry}

In this section, we provide the details of the calculation of the intrinsic metric tensor and the extrinsic curvature form for the gRNN memory manifold. We do not attempt to give a self-contained treatment of the elements of differential geometry, and refer to \cite{poisson2004relativist} for a concise reference of the necessary background material. 

\subsection{Intrinsic Geometry}

The intrinsic geometry is useful to describe the structure of the manifold from the perspective of observers moving on it. To begin, we introduce some notation we will use in this section. When splitting the population into active and frozen subsets, we can, without loss of generality, order the indices such that the first $N_{\rm a}$ neurons are active, which we denote by $h_{i}\equiv   a^{i}$, for $i = 1, ..., N_{\rm a}$ with a Latin subscript. The rest of the neurons are frozen, so that $h_{i} \equiv  f^{\mu}$, for $i = N_{\rm a} + 1, ..., N$, and $\mu = 1, .., i - N_{\rm a} , ..., N_{\rm f}$, which we denote with a Greek subscript. As before, we denote $\mu_{\sigma} = N_{\rm a}/N$ as the fraction of active neurons. The frozen activity $f^{\mu}$ is a natural choice of coordinates, since we determined before that there are exactly $N_{\rm f}$ linearly independent zero eigenvectors. The points which constitute the manifold at fixed $N_{\rm f}$ are determined by the fixed-point equations, 

\begin{align}
    G_{i}({\bf h}) = 0, \quad i = 1, ... , N_{\rm a} ,\label{eq:fp_constraint}
\end{align}

where in terms of ${\bf h} = ({\bf a}, {\bf f})$
\begin{align}\label{eq:fp_constraint_2}
    G_{i}({\bf a}, {\bf f}) = - a_{i} + J_{ij}^{\rm a} \phi(a^{j}) +J_{i\mu}^{\rm f} \phi(f^{\mu}).
\end{align}
where $J_{ij}^{\rm a} = J_{ij}$ for $i, j \le N_{\rm a}$, and $J_{i \mu }^{\rm f} = J_{i , N_{\rm a} + \mu}$ denote different blocks of the full connectivity matrix. 
The manifold geometry is determined locally by the set of normal and tangent vectors. Since we are interested in intrinstic geometry, we begin with the tangent vectors. These are given by
\begin{align}
    e_{\mu}^{i} = \frac{d h^{i}}{d f^{\mu}}, \quad \mu = 1, ..., N_{\rm f},
\end{align}

For $h$ satisfying the constraints (\ref{eq:fp_constraint}). Along the frozen directions, the tangent vector is aligned with the frozen directions $e_{\nu}^{\mu} = \delta_{\nu}^{\mu}$. Along the active directions, we define $e_{\mu}^{i} = d a^{i}/d f^{\mu} \equiv t_{\mu}^{i}$ for $i = 1, ..., N_{\rm a}$. By differentiating the constraint equations Eq. (\ref{eq:fp_constraint_2}), we may obtain a linear equation for $t_{\mu}^{i}$, 

\begin{align}
    t_{\mu}^{i} = \sum_{j} J_{ij}^{\rm a} \phi'(a^{j}) t_{\mu}^{j} +\sum_{\nu}J_{i\nu}^{\rm f} \phi'(f^{\nu}) \delta_{\mu}^{\nu}.
\end{align}

In matrix notation, this reads

\begin{align}
\underbrace{\left(-\mathbbm{I}+  J^{a} [\phi'({\bf \rm a})] \right)}_{\mathcal{D}_{\rm a}} {\bf t} = -J^{\rm f} [\phi'({\bf f})],
\end{align}
where we note that the active Jacobian $\mathcal{D}_{a}$ appears here on the LHS. This equation is easily solved to give the tangent vector

\begin{align}
t_{\mu}^{i} = \sum_{j = 1}^{N_{a}}\left( \mathcal{D}_{\rm a}^{-1}\right)_{ij} J_{j\mu}^{f} \phi'(f^{\mu}). \label{eq:tangent-vec}
\end{align}

To get the metric tensor, we use  $g_{\mu \nu} = \sum_{i} e_{\mu}^{i} e_{\nu}^{i}$ to get

\begin{align} \label{eq:app-metric-1}
    g_{\mu \nu} & =  \:  \delta_{\mu \nu} +   
                   \Gamma_{\mu \nu} \phi'(f_{\mu}) \phi'(f_{\nu})
\end{align}
where
\begin{align}
    \Gamma &= 
J^{{\rm f},T} \Sigma^{-1} J^{\rm f}, \quad {\rm and } \quad \Sigma = \mathcal{D}_{\rm a} \mathcal{D}_{\rm a}^{T}.\label{eq:app-metric-parts}
\end{align}

The matrix $\Gamma$ has the form of a Wishart matrix, and will have eigenvalues supported on a finite interval on the positive real line. It is interesting to note that the mean value is given by 

\begin{align}
   \frac{1}{N_{\rm f}} \langle {\rm Tr} \, \Gamma \rangle = \frac{\mu_{\sigma}}{1 - \mu_{\sigma} \langle (\phi')^{2}\rangle_{\rm a}} \equiv  \mu_{\sigma} \tau(\mathcal{D}_{\rm a}), \label{eq:moment_metric}
\end{align}
where $\langle (\phi')^{2} \rangle_{\rm a} = N_{\rm a}^{-1} \sum_{i = 1}^{N_{\rm a}} \phi'(a_{i})^{2}$, and  $\tau$ is the time constant defined in (\ref{eq:vRNN-spectral-abcissa-nonMFT}). A singularity in the metric would ordinarily signal a poor choice of coordinates. However, in the presence case, a singularity in the metric has a physical interpretation as signalling a point ($\tau = \infty$) where the attractor becomes unstable. Thus, our choice of coordinates $f^{\mu}$ is a good global choice to describe the attracting memory manifold.

We now proceed to prove Eq.(\ref{eq:moment_metric}). We assume that $N_{\rm a}$ scales with $N$ such that $\mu_{\sigma} = N_{\rm a}/N$ remains finite as $N \to \infty$. For a general $N_{\rm a}\times N_{\rm a}$ matrix $A$, we get to leading order in $N$,
\begin{align}
    \frac{1}{N_{\rm f}} \langle {\rm Tr}\left[ J^{{\rm f},T} A J^{\rm f} \right] \rangle_{J^{f}} 
    & =  \mu_{\sigma}\frac{1}{N_{a}}\operatorname{Tr} [A] ,\label{eq:moment_Jf}
\end{align}
Therefore, we are left to evaluate the mean trace norm of $\Sigma^{-1}$, which involves $J^{\rm a}$. From Eq.(\ref{eq:app-metric-parts}), using a series expansion for $\mathcal{D}_a^{-1}$, we get
\begin{align} \label{eq:app-metric-3}
   \Sigma^{-1} = \sum_{m,n = 0}^{\infty}  \left(
  [ \phi^{\prime} ] J^{{\rm a},T} 
    \right)^m
    \left(
J^{\rm a} [ \phi^{\prime} ]
    \right)^n.
\end{align}
Averaging over $J^{\rm a}$, only the ``non-crossing'' diagrams contribute to leading order in $N_{\rm a}$. These diagrams come from terms for which $m=n$. We thus get
\begin{align} \label{eq:app-metric-4}
    \frac{1}{N_{\rm a}} \langle  \operatorname{Tr}[\Sigma^{-1}] \rangle_{J^{\rm a}}
    = \left( 
      1 - \mu_{\sigma}\langle (\phi^{\prime})^2 \rangle_{\rm a}
      \right)^{-1}.
\end{align}
Combining this with Eq.(\ref{eq:moment_Jf}) then recovers Eq.(\ref{eq:moment_metric}).

\subsection{Extrinsic Geometry}

The metric tensor we have constructed is useful to characterise the intrinsic geometry. It tells us how much the state will change if we change the frozen coordinates. However, we are also interested in how the manifold is embedded in the full Euclidean space $\mathbbm{R}^{N}$. For instance, are the maps of the memory manifold more like the pages of an open book, or a collection of crumpled sheets of paper (it turns out the be the former, see Fig.(\ref{fig:lowD-manifold}). To determine this, we study the extrinsic curvature, or second fundamental form. The basic approach will involve tracking how the normal vectors evolve as we move along the manifold. If we perturb ${\bf h}$ by a vector field ${\bf h} + \epsilon {\bf v}$, then the constraints (\ref{eq:fp_constraint}) change

\begin{align}
   G_{i}({\bf h}') =  G_{i}({\bf h}) + \epsilon {\bf v} \cdot \frac{\partial G_{i}}{\partial {\bf h}}  + O(\epsilon^{2}).
\end{align}

If ${\bf v}$ is in the tangent space of the manifold at ${\bf h}$, then the second term vanishes. This indicates that the derivative of $G_{i}$ is normal to the surface. Furthermore, it indicates there are $N_{a}$ normal vectors, one for each constraint $G_{i}$. Unnormalized, they are 

\begin{align}
    \tilde{n}_{i}^{(j)} = \frac{\partial G_{j}}{\partial h^{i}} \equiv G_{j, i}.
\end{align}

To consider a general linear combination of these normal vectors, we construct the following potential function which vanishes on the memory manifold
\begin{align}
    \Phi({\bf h}) = \frac{1}{2}\sum_{i = 1}^{N_{a}} c_{i}  G_{i}^{2},
\end{align}
Here the coefficients $c_{i}$ are arbitrary. We denote the derivative by

\begin{align}
    \frac{\partial \Phi}{\partial h^{i}} = \Phi_{,i}.
\end{align}

The (normalized) normal vector is
\begin{align}
    n_{i} = \frac{\Phi_{,i}}{\sqrt{ g^{ij}\Phi_{,i} \Phi_{,j}}}.
\end{align}

Of course, the ambient geometry $\mathbbm{R}^{N}$ has $g^{ij} = \delta^{ij}$. Let us define
\begin{align}
\mathcal{D}_{ij} = \frac{\partial G_{i}}{\partial h_{j}} =  - \delta_{ij} + J_{ij} \phi'(h_{j}).
\end{align}

For the moment, there is no need to partition the populations into active/frozen. It will also make the notation simpler if we hold off on doing so. With the Jacobian, we can write the derivative of the potential as


\begin{align}
    \Phi_{,j} &=  \sum_{i  = 1}^{N_{a}} c_{i} G_{i}\mathcal{D}_{ij}.
\end{align}

We evaluate this on the manifold by setting $G_{k} = \epsilon$, and then taking the limit $\epsilon \to 0$. Defining the vector $\hat{l}_{i} = c_{i}$ for active indices, and $\hat{l}_{\mu} = 0$ for frozen indices, we get the normal vector

\begin{align}
    {\bf n}= \frac{\mathcal{D}^{T} \hat{l}}{\sqrt{\hat{l}^{T} \mathcal{D} \mathcal{D}^{T} \hat{l}}} .
\end{align}

With this construction, we can generate a family of $N_{a}$ linearly independent normal vectors by choosing $c_{k} = \delta_{ik}$. The gradient of the normal vector w.r.t. $h_{i}$ is found to be 

\begin{align}
    \frac{\partial n_{i}}{\partial h_{j}} 
    & = \kappa_{i} \delta_{ij} - n_{i} n_{j} \kappa_{j}, \label{eq:d_normal}
\end{align}
where we have defined a projected Hessian-like structure 
\begin{align}
   \kappa_{i} = \frac{1}{ \left( \hat{l}^{T} \mathcal{D} \mathcal{D}^{T} \hat{l}\right)^{1/2}}\sum_{k} \hat{l}_{k} J_{ki} \phi''(h_{i}) ,
\end{align}

and used the explicit form of the derivative of the Jacobian
\begin{align}
    \frac{\partial \mathcal{D}_{ki}}{\partial h_{j}} = J_{ki} \phi''(h_{i}) \delta_{ij}.
\end{align}


The next thing to do is project this derivative onto the manifold using the tangent vectors

\begin{align}
    e_{\mu}^{i}  = \frac{\partial h^{i}}{\partial f^{\mu}}.
\end{align}

Note that there will be $N_{\rm f}$ tangent vectors. Recall that the way to read these expressions is the following: $\mu$ labels the tangent vector, whereas $i$ labels the component of the tangent vector in $\mathbbm{R}^{N}$. We have that for all $\mu$, 

\begin{align}
   \sum_{i = 1}^{N} n_{i} e^{i}_{\mu} = 0.
\end{align}

Consequently, the second term in (\ref{eq:d_normal}) will vanish upon projection, and the extrinsic curvature, or second fundamental form, will be

\begin{align}
    K_{\mu \nu} &\equiv  \sum_{i} \kappa_{i} e_{\mu}^{i} e^{i}_{\nu},\\
    & = \sum_{i = 1}^{N_{\rm a}} \kappa_{i} \frac{\partial a^{i}}{\partial f^{\mu}} \frac{\partial a^{i}}{\partial f^{\nu}} + \sum_{\mu' = 1}^{N_{\rm f}} \kappa_{\mu'} \frac{\partial f^{\mu'}}{\partial f^{\mu}}, \frac{\partial f^{\mu'}}{\partial f^{\nu}} \\
    & = \phi'(f_{\mu}) \phi'(f_{\nu}) G_{\mu \nu}  + \kappa_{\mu} \delta_{\mu \nu},
\end{align}

where using Eq (\ref{eq:tangent-vec}) we find
\begin{align}
    G_{\mu \nu} = \sum_{i, j ,k= 1}^{N_{\rm a}}  \bar{J}^{\rm f}_{\mu j} \bar{\mathcal{D}}_{\rm a, ji}^{-1} \kappa_{i} \mathcal{D}_{\rm a,ik}^{-1} J^{\rm f}_{k \nu}.
\end{align}

or in matrix notation
\begin{align}
{\bf G} = \bar{J}^{\rm f} \bar{\mathcal{D}}_{\rm a}^{-1} [ \pmb{\kappa}] \mathcal{D}_{\rm a}^{-1} J^{\rm f}.
\end{align}
Here overbar indicates transpose, i.e. $A^{T} \equiv \bar{A}$. 
 The eigenvalues of the extrinsic curvature are the principal radii of curvature. As expected, the extrinsic curvature vanishes identically for piece-wise linear activation. Otherwise, we see that in the limit of large $N$, it actually vanishes on average 

\begin{align}
    \langle K_{\mu \nu}\rangle = 0,
\end{align}

where the expectation is taken over the quenched disorder $J$. This is ultimately due to the fact that $\kappa_{i}$ and $\kappa_{\mu}$ depend linearly on $J_{ij}$, which has zero mean. 

It is also interesting to study the second moment of $K$, which describes the variance of the principal curvatures. We find 
\begin{align}
\left\langle \operatorname{Tr}\left[ K^{2} \right]\right\rangle  &=  \frac{(1 - \mu_{\sigma})\langle \phi''^{2}\rangle_{\rm f}}{1 + \langle \phi'^{2}\rangle },\\
& + \frac{\left[ ( 1- \mu_{\sigma})\langle \phi'^{2}\rangle_{\rm f}\right]^{2} \left[  \mu_{\sigma} \langle (\phi'')^{2}\rangle_{\rm a} \right]}{\left(1 - \mu_{\sigma} \langle \phi'^{2}\rangle_{\rm a}\right)^{2}},\\
& + O(N^{-1}).
\end{align}

The interpretation of these results is as follows. We have computed the extrinsic curvature by first selecting a single direction normal to the memory manifold, as determined by the weight vector $\hat{l}$. The eigenvalues of the extrinsic curvature give us the $N_{\rm f}$ principal radii of curvature with respect to this normal vector. These principal curvatures tell us roughly how the normal vector varies as we move along the manifold. We find that on average, the principal radii are zero, and that for large $N$, the variance of the distribution of radii vanishes. Thus, for the gRNN, the embedding of the memory manifolds is statistically flat. 

We now proceed to prove the expression for the second moment. We start by writing it out explicitly

\begin{align}
\left\langle  {\rm Tr} [ K^{2}]\right\rangle  &=  \sum_{\mu, \nu} \phi_{\mu}'^{2}\phi_{\nu}'^{2} \left \langle G_{\mu \nu}^{2}\right\rangle, \label{eq:moment-1}\\
    &+\sum_{\mu} 2 \phi_{\mu}'^{2} \left\langle \kappa_{\mu} G_{\mu \mu}\right\rangle, \label{eq:moment-2} \\
    &+ \sum_{\mu} \left\langle \kappa_{\mu}^{2} \right\rangle \label{eq:moment-3},
\end{align}
where $\phi_{\mu}^{'} \equiv \phi'(f_{\mu})$. We attack this term by term. To facilitate some parts of the proof, we assume that $\hat{l}_{i} \sim \mathcal{N}(0, N_{\rm a}^{-1})$ are random Gaussian variables. Thus, we are studying the curvature with respect to a random normal vector.

First, we have for the projected Hessian variables the first two moments
\begin{align}
\langle  \kappa_{i}\rangle &\approx \frac{1}{\sqrt{\mathcal{N}}} \left\langle  \phi''(h_{i}) \hat{l}_{k} J_{k \mu} \right\rangle = 0, \label{eq:mean_kappa}\\
\left\langle  \kappa_{i} \kappa_{j} \right\rangle &\approx  \delta_{ij} \frac{1}{\mathcal{N}} \frac{1}{N} (\phi''(h_{i}))^{2},\label{eq:variance_kappa}
\end{align}
where we have taken the expectation values over $\hat{l}$ and $J$. We have also approximated the normalization factor by its expected value, which we expect to be valid in the limit of large $N$, 
\begin{align}
\mathcal{N} \equiv \langle \hat{l}^{T} \mathcal{D} \mathcal{D}^{T} \hat{l} \rangle , 
& \approx  1 + \frac{1}{N} \sum_{i = 1}^{N} \phi'(h_{i})^{2}.
\end{align}

These results determine the last term Eq. (\ref{eq:moment-3})

\begin{align}
\sum_{\mu} \left\langle \kappa_{\mu}^{2} \right\rangle &= \frac{1}{\mathcal{N}} \frac{1}{N} \,  \sum_{\mu} \left(\phi''(f_{\mu})\right)^{2},\\
& = (1- \mu_{\sigma})\frac{\langle \phi''^{2}\rangle_{\rm f}}{1 + \langle \phi'^{2}\rangle },
\end{align}
where we have introduced $\langle (\phi'')^{2}\rangle_{\rm f} \equiv N_{\rm f}^{-1} \sum_{\mu}( \phi''(f_{\mu}))^{2}$. 

 Next, we note that $\kappa_{\mu}$ will depend linearly on $J^{\rm f}$, whereas $G_{\mu \nu}$ depends on $\kappa_{i}$ which scales with $J^{\rm a}$. For this reason, the second line (\ref{eq:moment-2}) will vanish, since it involves an odd number of factors of $J^{\rm f}$, which is a centered Gaussian variable.

 Finally, defining $M = \bar{\mathcal{D}}_{\rm a}^{-1} \left[ \pmb{\kappa}\right] \mathcal{D}_{\rm a}^{-1}$, we evaluate the moment in the first line (\ref{eq:moment-1}) by first averaging over $J^{\rm f}$

\begin{align*}
 \left\langle G_{\mu \nu}^{2} \right\rangle = \frac{1}{N^{2}} \langle {\rm Tr} \, M^{2} \rangle + \frac{1}{N^{2}} \delta_{\mu \nu} \left \langle  \left( {\rm Tr} \, M \right)^{2} + {\rm Tr} \, M^{2}\right \rangle  + O(N^{-1}).
\end{align*}
The first term gives the leading order contribution after summing over $\mu, \nu$. To evaluate it, we utilize Isserliss' or Wick's theorem for expectations of Gaussian random variables. We first take the contraction over the Gaussian random variables $\hat{l}$ and $J^{\rm a}$ appearing in $\kappa_{i}$, then contract over the remaining $J^{\rm a}$ . At each step, we keep only the leading terms in $N$. The result is

\begin{align}
  \left\langle {\rm Tr} [ M^{2}] \right\rangle = \mu_{\sigma} \frac{\langle (\phi'')^{2}\rangle_{\rm a}}{\left(1 - \mu_{\sigma} \langle \phi'^{2}\rangle_{\rm a}\right)^{2}} + O(N^{-1}).
\end{align}

Using this, we may evaluate the first contribution to the second moment (\ref{eq:moment-1}) as

\begin{align}
   \sum_{\mu, \nu} \phi_{\mu}'^{2}\phi_{\nu}'^{2} \langle G_{\mu \nu}^{2}\rangle 
   & = \frac{\left[ ( 1- \mu_{\sigma})\langle \phi'^{2}\rangle_{\rm f}\right]^{2} \left[  \mu_{\sigma} \langle (\phi'')^{2}\rangle_{\rm a} \right]}{\left(1 - \mu_{\sigma} \langle \phi'^{2}\rangle_{\rm a}\right)^{2}},
\end{align}

which concludes the proof.

\section{Details of the numerics for global geometry} \label{app-sub-globalgeometry}

Here we provide the details for generating Fig. \ref{fig:lowD-manifold}. We use a gRNN with the following parameters: $N=3,g=2.0$ and the following randomly generated connectivity matrices:
\begin{align}
    J = & \left( \begin{array}{ccc}
-0.14  &  0.13  &  -0.62\\
 -0.45 &  -0.19 &   -1.50\\
 0.27 & -0.28 & -1.16
 \end{array}\right), \\
 W = & \left( \begin{array}{ccc}
0.57  &  -0.26  &  0.95\\
 0.02 &  0.27 &   -0.46\\
 1.0 & -0.04 & -0.04
 \end{array}\right).
\end{align}

The initial conditions are sampled from a grid in the cube $[-1,1]^3$, and the dynamics are run till a fixed-point is reached. The final state is plotted, with a color indicating which subset of the 3 neurons are part of the frozen population.

\section{Frozen-stabilisation in neurons with ``notch'' gating}\label{app:notch-gating}

In the main text, we considered gating functions with threshold behaviour that allowed neurons to be frozen over a relatively large range of the phase space. For instance, when the gating function involved a step function, the neuron was frozen whenever the argument to the step function was negative. One possible concern is whether the integrator functionality of the network depends on gating function to freeze the neurons over a wide range of inputs? 

Here, we address this concern to show that even when the individual neurons can  only function as integrators over a small range of inputs, frozen-stabilisation permits the full network to integrate inputs over a much larger range -- i.e. the integrator functionality emerges out of the collective interaction between the neurons and is not solely a function of the properties of a single neuron. To do this, we consider neurons with ``notched'' gating functions. Let us again consider the dynamics prescribed in eq. \ref{eq:large-gRNN}, with a modified gating function $\sigma_{notch}$

\begin{align} \label{eq:notch-gRNN-1}
    \frac{d \mathbf{h}}{d t} = \: &\sigma_{notch}(W \mathbf{h}) \odot[-\mathbf{h}+J \phi(\mathbf{h})] \\
    \sigma_{notch}(x) =& 
    \left\{
\begin{array}{ll}
      0 & 0 \leq |x| \leq \beta/2 \\
      1 & \textrm{o/w}\\
\end{array} 
\right. 
\end{align}

Thus, the individual neurons can only integrate inputs over a range $\beta/2$. However, as show in Fig. \ref{fig:notch-gating}, we see that a network of such neurons with the ``notched'' gating is able to integrate over a larger range of inputs, thus demonstrating that the integrator functionality is indeed an emergent property of the network dynamics and not simply a reflection of the capabilities of single neurons.


\begin{figure}
\begin{centering}
\includegraphics[scale=0.4
]{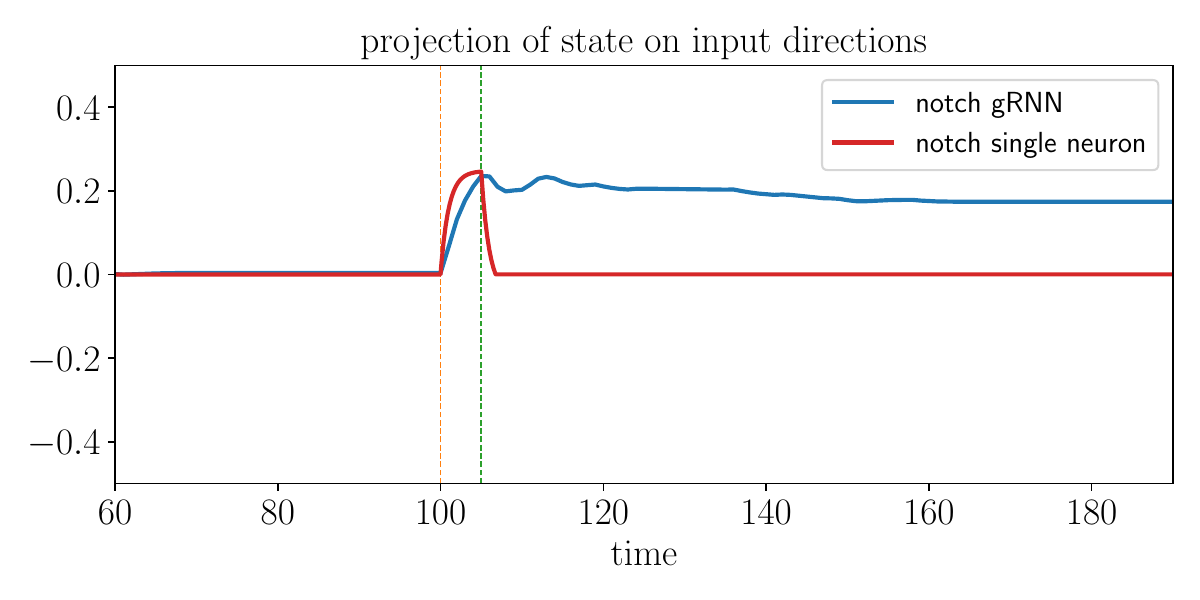}
\par\end{centering}
\caption{\label{fig:notch-gating} {\it Integrator function of the gRNN with ``notch'' gating}  When the neurons have a notched gating function as described in eq. \ref{eq:notch-gRNN-1}, each neuron is a poor integrator, but as shown in the figure, frozen-stabilisation allows the network to function as a reasonably good integrator. Inputs were applied between the dashed vertical lines, and as can be seen the individual neurons do not maintain a memory trace of the inputs, whereas the full network is able to integrate the inputs and maintain it in memory long after the inputs are removed. For this figure, the width of the notch was chosen to be $\beta = 0.1$, and the network size $N=100$.}
\end{figure}


\section{Implementation in Balanced Networks}\label{app-balance}

We point out in the main text that frozen stabilisation arises reliably in networks with gates that dynamically modulate the time constant, as in Eq. \ref{eq:large-gRNN}. In this appendix, we show that such a state-dependent time constant should also arise in nonlinear balanced networks by a mechanism known as negative derivative feedback. 

In \cite{lim2013balanced} a mechanism is proposed for producing negative-derivative feedback that promotes enhanced lifetimes in a neural circuit. In this appendix, we point out that gating can be viewed as such a feedback. The observation is very simple. We emphasized in the main text that if the discreteness of the gate is relaxed, one still observes (approximate) frozen stabilisation so long as there is a clear separation of timescales. This allows us to perform the crucial trick of writing the gating function in Eq. \ref{eq:large-gRNN} as a state-dependent time constant, i.e. $\sigma({\bf h}) = 1/\tau({\bf h})$, where $0 <\tau({\bf h})<\infty$. The dynamical equations can then be written


\begin{align}
    \tau_{i}({\bf h}) \frac{d h_{i}}{dt} = G_{i}({\bf h}).
\end{align}

If we let $\tau({\bf h}) = \tau + \gamma({\bf h})$ for $\gamma >0$ and a constant $\tau$, we have the effective description

\begin{align}
    \tau \dot{{\bf h}} = - \gamma({\bf h}) \dot{\bf h} + {\bf G}({\bf h}). \label{eq:damping_gate}
\end{align}

The first term on the rhs of Eq.\ref{eq:damping_gate} is precisely ``negative-derivative" feedback. In \cite{lim2013balanced}, the damping coefficient $\gamma$ does not explicitly depend on the state ${\bf h}$. However, following their  set-up, we find that introducing a nonlinearity is enough to produce nontrivial state-dependence in the damping coefficient for the simple balanced network they considered. This can be seen starting from their system of equations
\begin{align}
    \tau \dot{r} &= - r + J s_{E} - J s_{I} + I,\\
    \tau_{E}\dot{s}_{E} & = - s_{E} + \phi(r),\\
    \tau_{I} \dot{s}_{I} & = - s_{I} + \phi(r).
\end{align}

Here, $r$ represents the firing rate, $s_{E/I}$ are the excitatory/inhibitory recurrent feedbacks, and $I$ is the external input. A crucial assumption for the derivation is that the strength of feedback $J$ is taken to be the same for both recurrent inputs, while the intrinsic time constants $\tau_{E/I}$ are different. Expanding the Laplace transform of the difference $s_{E} - s_{I}$ for small frequencies $|u \tau_{I}|<<1$ produces 


\begin{align}
    L[s_{E} - s_{I}](u) &\approx (\tau_{I} - \tau_{E}) u L[ \phi(r)] \\
    &= - \Delta \tau  L \left[ \frac{d \phi(r)}{dt}\right], \quad \Delta \tau = \tau_{E} - \tau_{I}.
\end{align}
This expansion is valid in a slowly-varying regime where $|u \tau_{I}| <<1$. Inserting this into the firing rate equation gives

\begin{align}
    \tau \dot{r} & \approx  - r - \Big( J \Delta \tau  \phi'(r) \Big) \dot{r} + I,
\end{align}
from which we can identify $\gamma(r) \equiv J \Delta \tau \phi'(r)$ by inspection. This simple example demonstrates that state-dependence of the effective damping coefficient should be a generic phenomena, and appears to be tied to nonlinearities. Since a state-dependent damping constant is equivalent to timescale gating via Eq.\ref{eq:damping_gate}, we have shown that balanced networks with nonlinearities can give rise to an effective dynamical gating of time constants. 

\end{widetext}

\end{document}